\documentclass[12pt,preprint2]{aastex}
\usepackage{color}
\usepackage{bm}

\slugcomment{\footnotesize {\underline {Preprint: accepted 2/15/2016 for publication in {\it Ap. J.}}}}

\shorttitle{Jupiter Gravity and Interior}

\shortauthors{Hubbard and Militzer}

\begin{document}

\title{A Preliminary Jupiter Model}

\author{W. B. Hubbard}
\affil{Lunar and Planetary Laboratory, The University of Arizona, Tucson, AZ 85721, USA.}

\and

\author{B. Militzer}
\affil{Department of Earth and Planetary Science, Department of Astronomy, University of California, Berkeley, CA 94720, USA.}

\begin{abstract}

In anticipation of new observational results for Jupiter's
axial moment of inertia and gravitational zonal harmonic coefficients
from the forthcoming {\it Juno} orbiter, we present a number of
preliminary Jupiter interior models.
We combine results from {\it ab initio} computer simulations of
hydrogen-helium mixtures, including immiscibility calculations, with
a new nonperturbative calculation of Jupiter's zonal harmonic
coefficients, to derive a self-consistent model for the planet's
external gravity and moment of inertia. We assume helium rain
modified the interior temperature and composition profiles. Our
calculation predicts zonal harmonic values to which measurements
can be compared. Although some models fit the observed (pre-{\it Juno})
second- and fourth-order zonal harmonics to within their error bars, our
preferred reference model predicts a fourth-order zonal harmonic whose
absolute value lies above the pre-{\it Juno} error bars.
This model has a dense core of about 12 Earth masses, and a
hydrogen-helium-rich envelope with approximately 3 times solar
metallicity. 
\end{abstract}


\section{Introduction}

In July 2016, the $\it {Juno}$ spacecraft will enter a bound orbit around Jupiter,
and then complete $\sim 30$ further low-periapse orbits over a period of approximately
one year.  Measurements of the spacecraft's accelerations may reach a precision of $\sim 1 \phn \mu$gal \citep{kas10}, allowing determination of Jupiter's external gravitational potential, $V$,
to a relative precision approaching $\sim 10^{-9}$.  In roughly the same time frame,
the $\it {Cassini}$ spacecraft will execute $\sim$ 22 low-periapse orbits around Saturn, making similar measurements of Saturn's external gravity potential.  The nonspherical components of $V$ provide
information about a planet's interior mass distribution.

In this paper, we construct static interior models
intended to represent the present state of Jupiter,
using a pressure-density relation
$P(\rho)$ derived
from DFT-MD theory for the equation of state of the primary constituent of Jupiter and Saturn,
a mixture of hydrogen and
helium; see \citet{MH2013} and \citet{Militzer2013}.
This barotrope is
used to calculate the zonal harmonic coefficients $J_{2n}$,
making various assumptions about the interior temperature distribution and core mass.
Physically-motivated adjustments of the barotrope are
made to achieve agreement with the observed $J_2$ (Table 1) and discrepancies
with currently-observed higher $J_{2n}$ are discussed.
Lines 2-11 of Table 1 give calculated values from interior models discussed in Section 4.

To obtain a barotrope, we start with the grid of {\it ab initio}
adiabats derived in \citet{Militzer2013} and \citet{MH2013}. These
adiabats were determined with density functional molecular dynamics
(DFT-MD) simulations using the Perdew-Burke-Ernzerhof (PBE)
functional~\citep{PBE} in combination with a thermodynamic integration
(TDI) technique to determine the full, nonideal entropy. The
simulation cells contained a mixture of $N_{\rm He}=$18 helium and
$N_{\rm H}=220$ hydrogen atoms, corresponding to a helium mass
fraction of $Y$=0.245, close to the solar value.  As discussed in
\citet{MH2013}, each adiabat is characterized by the value of its
absolute entropy per electron, $S/k_B/N_e$, where $k_B$ is Boltzmann's
constant and $N_e$ is the number of electrons.  Hereafter we denote
this quantity with the simpler symbol $S$.

Recently \citet{becker2014} constructed Jupiter models based on
equations of state that were also derived with DFT-MD simulations but
their approach differs in two respects. Becker {\it et al.} performed
simulations for hydrogen and helium separately and then envoked the
ideal mixing assumption while we simulated an interacting
hydrogen-helium mixture directly. While we computed the full, nonideal
entropy with TDI, Becker {\it el at.} obtained the entropy indirectly
by fitting the internal energy and pressure, which are available in
standard DFT-MD simulations. \citet{becker2014} reported deviations
between 4 and 9\% when they compared their EOS with \citet{MH2013}.
Such deviations could have a significant repercussion on values of
zonal harmonics for interior models.

In this paper, we
  use the term ``entropy'' and the symbol $S$ as a proxy for an
  adiabatic temperature $T$ vs. pressure $P$ relation for the
  fixed-composition mixture of H and He only (He mass fraction
  $Y_0=0.245$), as determined by our detailed DFT-MD simulations.  The
  simulations give the absolute entropy and other dependent variables
  as a function of $T$ and $P$, for this specific composition.  As
  discussed in Section \ref{comppert} below, for the purpose of
  calculating general pressure-density relations, the same $T(P)$
  relation is taken to apply to adiabats with small, constant
  perturbations to the composition of the simulations.
Moreover, the $S$ of the outermost layers of the model is
determined by requiring a match to the Galileo Probe measurements of
$T(P)$; see Figure~\ref{TvsP_gap}.  The corresponding adiabat from
our simulations has $S = 7.08$.  Now, if we perturb this composition by
changing $Y$ and increasing $Z$, how might the adiabatic $T(P)$ change,
for $P \textgreater 20$ bar, and how might this affect the barotrope?
Let the Gr{\"u}neisen parameter
$\gamma = (\rho/T) (\partial T /\partial \rho)_S$, where $\rho$ is the mass
density.  Suppose we have a compositional perturbation to $Y$ and/or $Z$,
of the order of $\sim 0.01$.  This might lead to a perturbation
$\Delta \gamma \sim 0.01$.  Over a density range of three orders of magnitude,
roughly spanning the jovian mantle, this value $\Delta \gamma$ would imply
a cumulative change of temperature of $\sim 7\%$, with respect to the baseline
$T(P)$.
According Mie-Gr{\"u}neisen theory \citep{ZT1978}, the
  thermal pressure makes up only 10\% of the total pressure in the
  relevant Jupiter layers.  Therefore, we expect the fractional change
  in density to be on the order of $\sim 0.1 \times 0.01 = 0.001$.
  This amount is so small that it is unlikely to affect any of our
  model predictions. It is certainly smaller than the
  previously-mentioned 4 to 9\% discrepancy with \citet{becker2014}.

Our {\it ab initio} calculations show that under
  jovian interior conditions, there is no distinct phase transition
  from molecular (diatomic, insulating) hydrogen to metallic
  (monatomic, conducting) hydrogen~\citep{Vo07}.  However, for
convenience in this paper, the term ``molecular'' layer means layers
at pressures below 1 Mbar, where the hydrogen is mostly diatomic.
Likewise, the term ``metallic'' layer means layers at pressures above
$\sim 2$ Mbar, but still external to a central dense core.

By combining our {\it ab initio} calculations for Jupiter's interior
adiabat \citep{Militzer2013} with the {\it ab initio} hydrogen-helium
immiscibility calculations by ~\citet{Mo2013}, we predict that helium
rain occurs in Jupiter's interior. While the detailed physics and
dynamics of helium rain is not yet understood, we make the assumption
that this process introduces a superadiabatic temperature gradient and
a compositional difference between the outer, molecular layer and
inner, metallic layer. In our models, the $T(P)$ of the molecular
layer is set by the measurements of the Galileo entry probe while the
$T(P)$ of the metallic layer is a free parameter that we can adjust
between two limits. The value of $S$
labeling $T(P)$ for the metallic layer cannot be too high
  because otherwise no helium rain would have occurred in Jupiter
  according to DFT-MD simulations. The value of $S$ labeling
 $T(P)$ cannot be below the
  Galileo value because, we assume, the cooling of the metallic
  layer is less efficient.  The assumption of reduced cooling
of the metallic layer is consistent with specific
  models constructed by \citet{nettelmann2015} who studied the
  evolution of jovian interior temperature profiles under the
  influence of H/He demixing and layered double diffusive convection
  (see upper left-hand panel of Figure 10 of that paper).

For the molecular layer, we assume the helium abundance that was
measured by the Galileo probe \citep{vonzahn-jgr-98, mahaffy-jgr-2000}. We derive the helium contents in the
metallic layer by assuming the planet as a whole has a protosolar helium
abundance \citep{Lodders03}. The distribution of heavier elements throughout the planet
is not well understood. The capture of comets has enriched the
envelope over time. Similarly the erosion of the core may have added
icy and rocky materials to the envelope~\citep{WilsonMilitzer2012,WilsonMilitzer2012b,Wahl2013,SiO2}. Given these uncertainties, we
introduce three model parameters: the mass of today's dense core, the
heavy element (``metals'') mass fraction in the molecular layer and that in the metallic
layer. We assume that both layers are homogeneous and interpolate
between both compositions to derive an estimate for the structure of the helium rain
layer. Model predictions are not sensitive to details of this procedure because, in
Jupiter, the interpolation layer between 1 and 2 Mbar contains very little mass.

This article is organized as follows. In Section 2, we
  describe how we deal with hydrogen-helium immiscibility. In Section
  3, we discuss how one perturbs the helium abundance in a particular
  EOS and how heavy elements are introduced. In Section 4, we discuss
  the EOS of different planetary ices and present results from
  additional {\it ab initio} simulations. In Section 5, we introduce
  our reference Jupiter model and discuss variations from it. Before
  we conclude, we describe in Section 6 how the moment of inertia is
  derived from CMS theory. In the Appendix, we provide additional
  details about the CMS calculations.

\section{Adiabats and Hydrogen-Helium Immiscibility} \label{adimm}

\begin{figure}
\epsscale{1.0}
\plotone{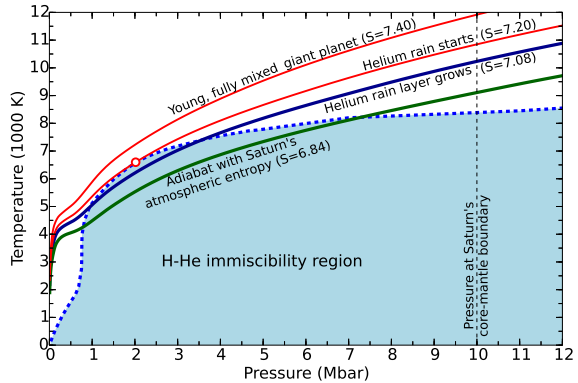}

\caption{General view of effect of immiscibility on Jupiter (and Saturn)
evolution.  The top two curves are DFT-MD adiabats with relatively high entropy per electron.  The adiabat for $S \approx$ 7.20 osculates the boundary of
the region of H-He immiscibility, while the adiabat just below it
has $S =$ 7.08,
which yields a temperature vs. pressure relation in the Jovian troposphere that matches
corresponding data from the Galileo Probe \citep{seiff-1998}. 
The lowest adiabat has $S =$ 6.84, which yields a
temperature vs. pressure relation that roughly matches Saturn's
tropospheric profile \citep{Lindal-1985}. The pressure at Jupiter's
core-mantle boundary (about 40 Mbar) is not shown on this figure.
\label{phase_bdry2}}
\end{figure}

\begin{figure}
\epsscale{1.0}
\plotone{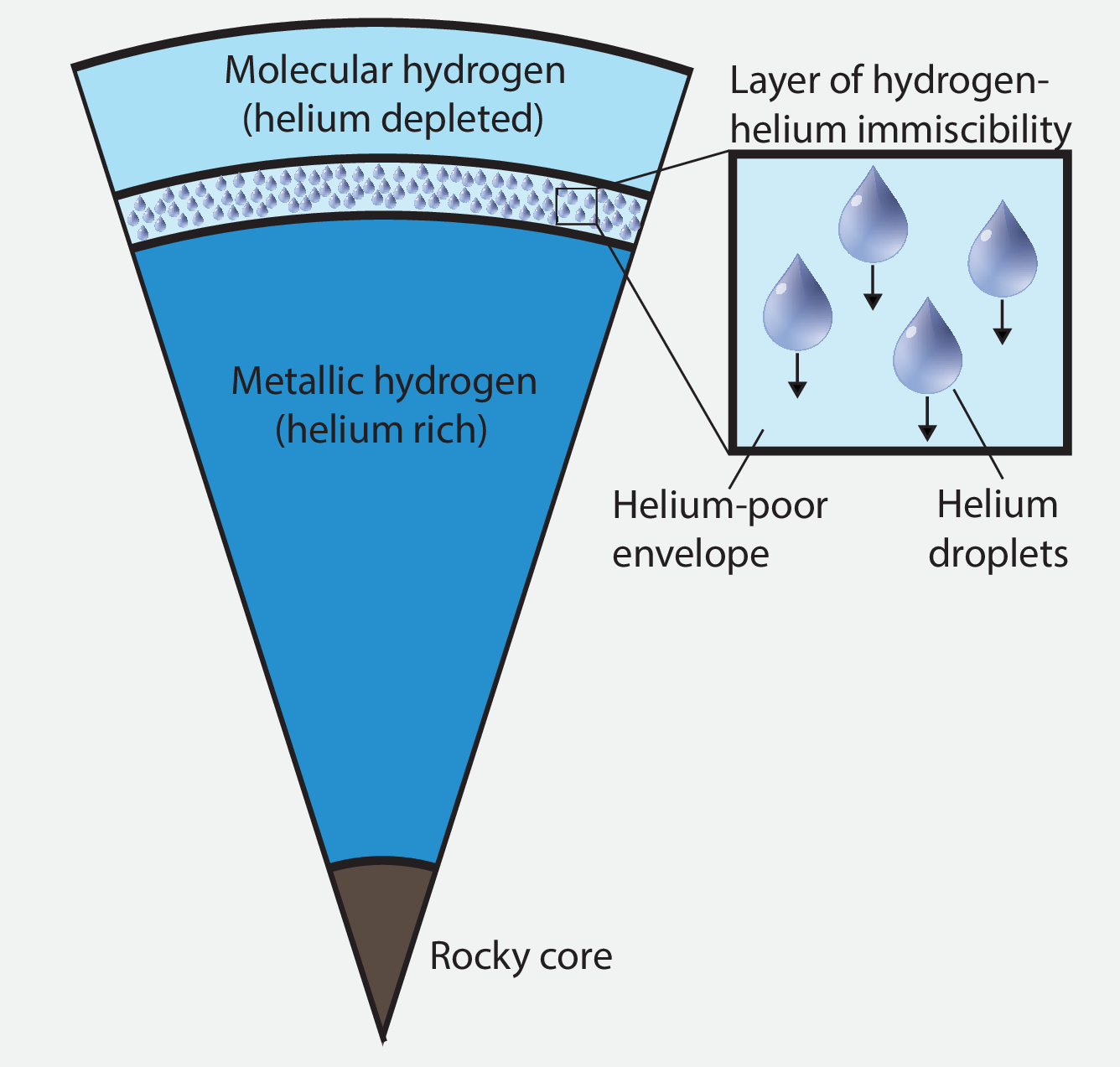}
\caption{Diagram showing the location of
the hydrogen-helium immiscibility layer in Jupiter.
\label{rain}}
\end{figure}

\begin{figure}
\epsscale{1.0}
\plotone{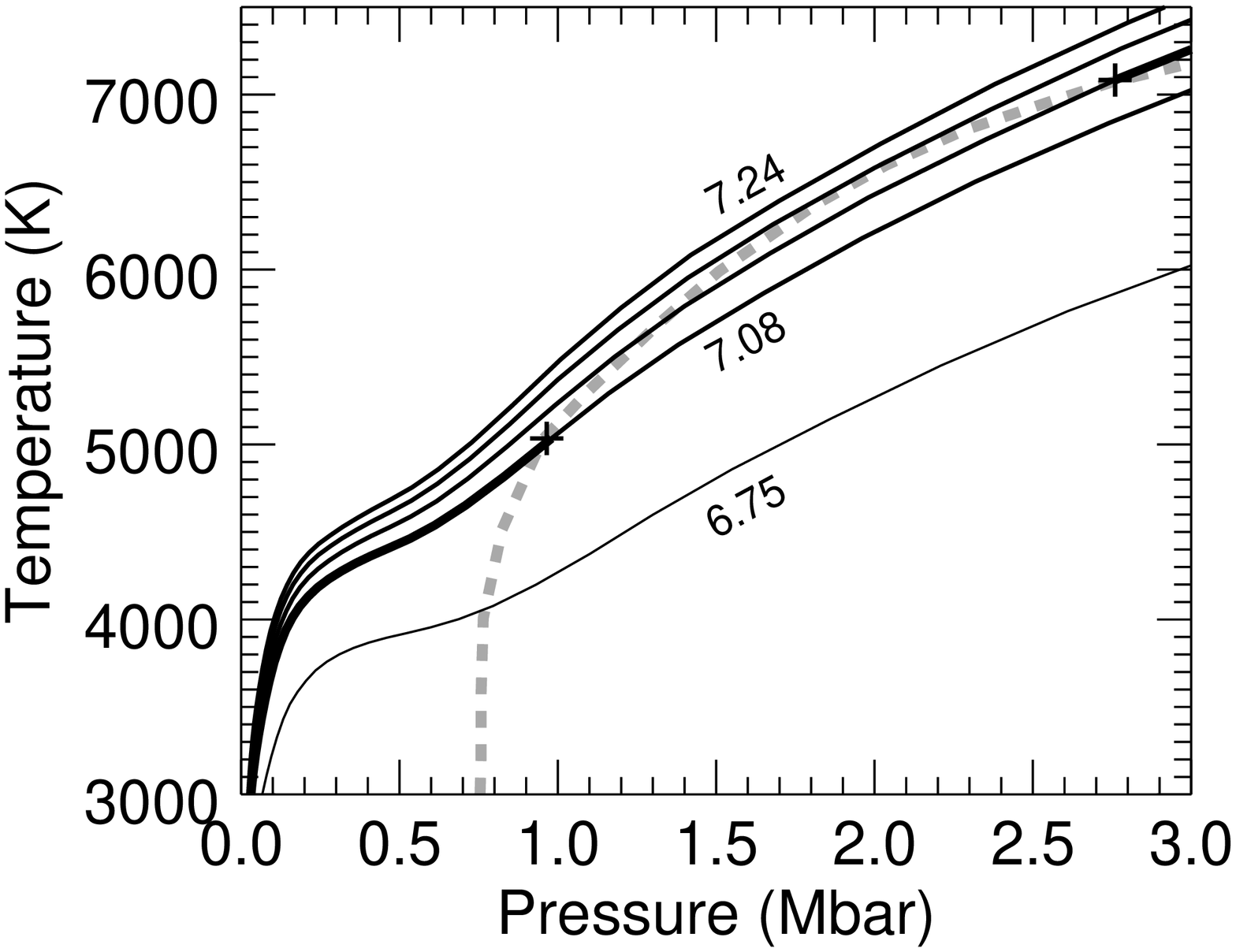}
\caption{Temperature-pressure relations used in the models.  DFT-MD adiabats are labeled with their entropy
per electron $S  = $
7.24 (top) to 6.75 (bottom).  The two middle (unlabeled) adiabats have
$S =$ 7.20 and 7.13.  The preferred temperature-pressure
relation of this paper is shown as a heavier curve following the
Galileo Probe adiabat to the immiscibility boundary \citep{Mo2013}
shown with a dashed curve.  At pressures higher than 2.7 Mbar we assume a higher-entropy
adiabat with $S =$ 7.13 (heavier curve).
\label{phase_bdry4}}
\end{figure}

Figure~\ref{phase_bdry2} shows a plot of temperature, $T$, versus
pressure, $P$, for a family of such adiabats as well as the
hydrogen-helium immiscibility domain derived from {\it ab initio}
simulations by~\citet{Mo2013}. These simulations also used the DFT-MD
technique in combination with the PBE functional and TDI method to
compute the entropy. They are thus fully compatible with the abiabats
from \citet{MH2013}. As is evident in Figure~\ref{phase_bdry2}, the
interiors of both Jupiter and Saturn enter a region at a pressure
$\sim$ 1 Mbar where helium in solar proportion to hydrogen becomes
immiscible. Both planets are thus are likely to have layers with
helium rain. Figure~\ref{rain} depicts the location of this layer in
Jupiter. This prediction is a direct consequence of combining the {\it
  ab initio} immiscibility and adiabat calculations with measurements
of the planets' tropospheric $T$ vs. $P$ profiles \citep{seiff-1998,
  Lindal-1985}.  No temperature or pressure adjustments of the
immiscibility domain were needed.

  Heavy elements were not considered in this analysis. Depending on
  the concentration, a small correction to the adiabatic profile would
  be plausible. We also note that \citet{Mo2013} performed the
  immiscibility calculations for $Y=0.25$ which differs slightly
  from the protosolar value. However, this concentration difference
  does not change the immiscibility temperature to a significant
  degree. Based on the analysis in ~\citet{nettelmann2015}, we
  estimate this correction to be of the order 160~K only.

The hydrogen-helium immiscibility hypothesis was first invoked to
explain Saturn's luminosity excess~\citep{Stevenson77a,Stevenson77b}.
In the immiscibility layer, helium droplets would form and 
rain down into the deeper interior, resulting in a gradual removal of
helium from the planets outer layer. The associate release of
gravitational energy provides an energy source to explain Saturn's
luminosity excess.

Whether helium rain occurs on Jupiter is less certain. Its interior is
hotter and no helium rain is need to explain its present
luminosity~\citep{FH04}. The Galileo entry probe measured a small
helium depletion in Jupiter's upper atmosphere (0.234 by mass compared
to 0.274, the protosolar value~\citep{Lodders03}. Perhaps the strongest
evidence for helium rain to occur on this planet comes from the
depletion of neon. The Galileo measurements showed that there is ten times
less neon in Jupiter's atmosphere compared to solar values.
\citet{WilsonMilitzer2010} demonstrated with {\it ab initio}
simulations that neon has a strong preference for dissolving in the
forming helium droplets. This offered an explanation for the neon
depletion and provided strong, though indirect evidence for helium
rain to occur on Jupiter.

According to the more recent {\it ab initio} calculations, the present
Jupiter would encounter the immiscibility domain at pressures above
$\sim 0.9$ Mbar (Figure~\ref{phase_bdry2}). In Saturn, the domain is
entered at $P \sim 0.8$ Mbar. If a cooling scenario for Jupiter or
Saturn involves a steady decrease of entropy with time, then the onset
of helium rain would occur when the interior adiabat first touches the
boundary of the helium immiscibility domain.  The curvature of the
boundary is such that a H-He adiabat with $S \approx$ 7.20
osculates the boundary at $P \sim 2$ Mbar and $T \sim 6600$ K.

We are thus faced with the task of deriving a barotrope $P(\rho)$ for
present-day Jupiter which is consistent with the properties of dense,
hot hydrogen-helium mixtures shown in Figure~\ref{phase_bdry2} and
with Jupiter's presumed cooling history.  A detailed, dynamical
calculation of the process of helium rain and subsequent evolution of
Jupiter's interior temperature profile is beyond the scope of the
present paper, whose aim is to infer a jovian barotrope based on
current knowledge of Jupiter's composition and thermal state, and on
current results from {\it ab initio} simulations of hydrogen-helium
mixtures at high pressure. The resulting barotrope is used here to
predict Jupiter's higher zonal harmonic coefficients, whose values are
to be measured by {\it Juno}.

\begin{figure}
\epsscale{1.0}
\plotone{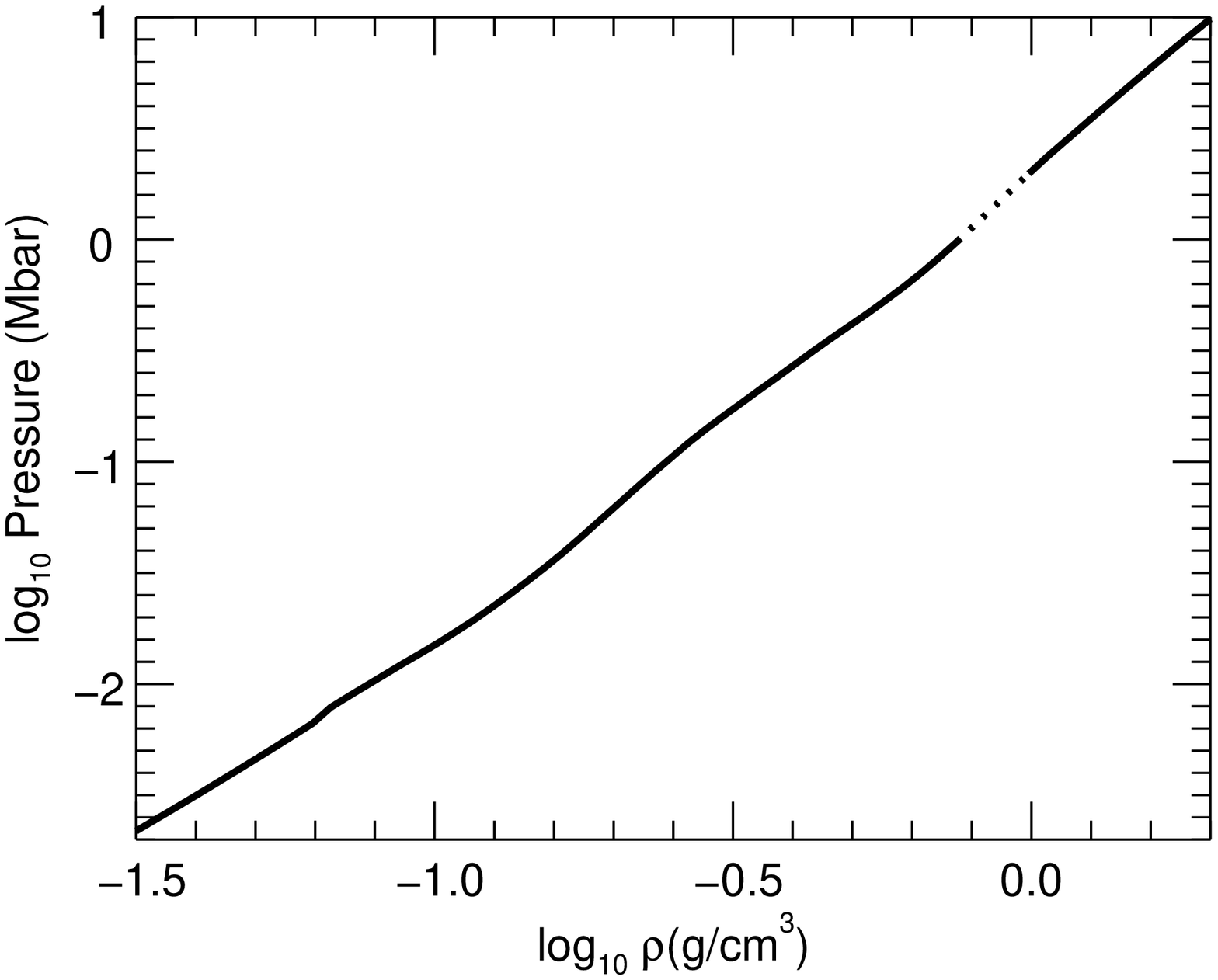}

\caption{The initial approximation for the present-Jupiter
barotrope; the abscissa is $\rho_0(P)$.
The gap corresponds to the region between the two
plus symbols in Figure~\ref{phase_bdry4}.  To the left of the
gap the entropy is $S =$ 7.08, while to the right
$S =$ 7.13.  Both adiabats are for constant $Y_0$=0.245.
Since we do not have DFT-MD simulation
data at very low densities, we switch back to the SC model below
0.0670$\,$g$\,$cm$^{-3}$, where a small (and unimportant)
density discontinuity $\sim 2\%$ can be seen.
\label{Pvsrho_gap}}

\end{figure}

\begin{figure}
\epsscale{1.0}
\plotone{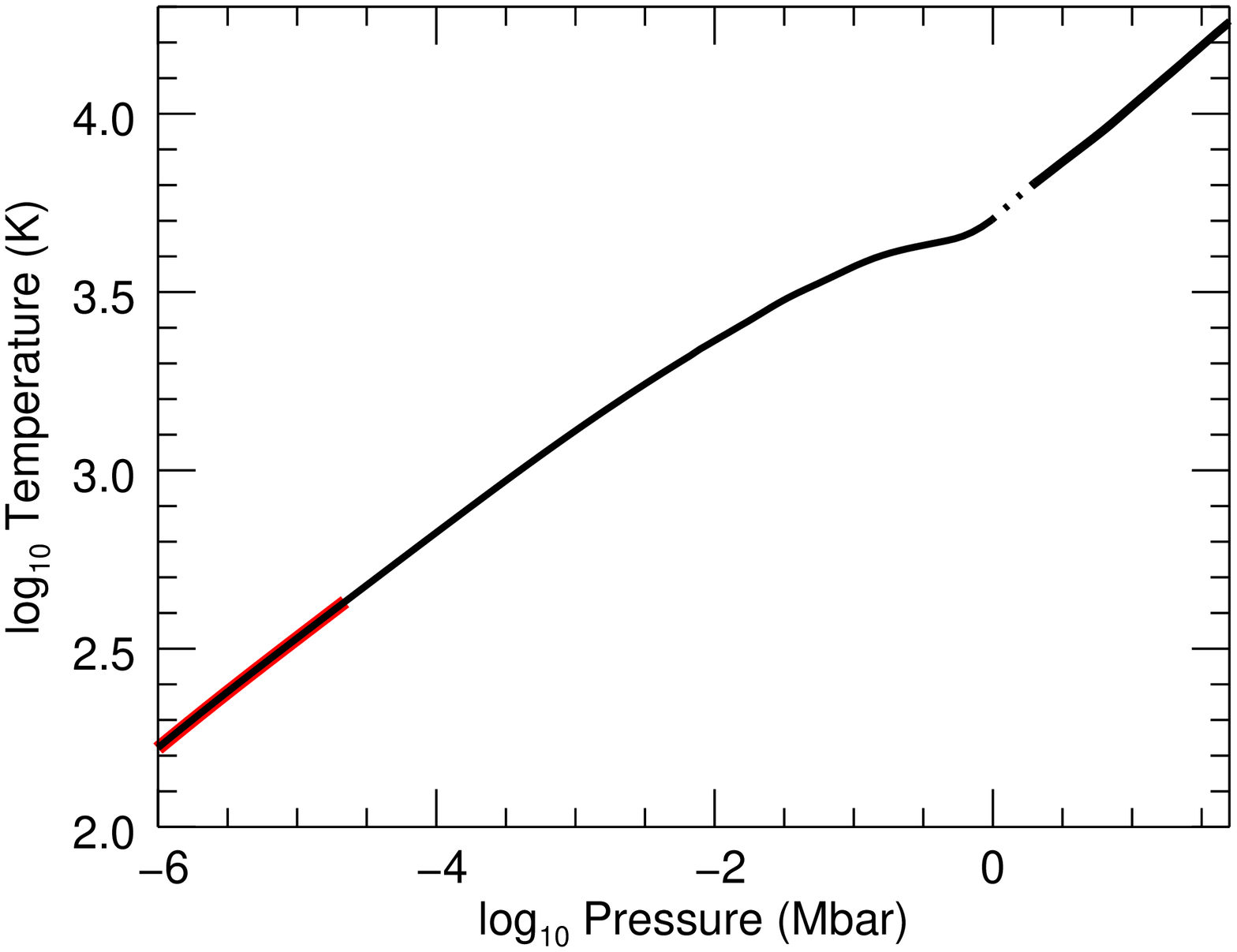}
\caption{The $T$ vs. $P$ relation for the two adiabats shown in
Figure~\ref{Pvsrho_gap}.  Thick curve up to $P = 22$ bar shows Galileo Probe measurements.
The $S =$ 7.08 adiabat's $T$ vs. $P$ relation matches
Galileo Probe data.
\label{TvsP_gap}}
\end{figure}

Thus, we make the simplifying assumption that the cooling of early
Jupiter to an interior adiabat $S \approx$ 7.20, corresponding the
onset of immiscibility, then leads to reduced heat transport in the
region around $P \ge 2$ Mbar, effectively slowing interior temperature
decline, while layers at lower pressures continue to transfer heat to
Jupiter's atmosphere.  In this scenario, the present-day Jupiter
barotrope for pressures $\le 1$ Mbar lies on the Galileo Probe adiabat
with reduced He abundance, but at somewhat higher pressures,
temperatures follow a higher-entropy adiabat with a slightly-above
protosolar helium abundance, $Y=0.28$.
The interior adiabat would be expected to lie between $S
\approx$ 7.20 (for no heat transport across the immiscibility region)
and $S \approx$ 7.08 (for efficient heat transport across the
immiscibility region).  In the study that we present here, our preferred
model has an interior adiabat with $S =$ 7.13 (shown
as a heavy line in Figure~\ref{phase_bdry4}).  We refer to this model
as Model DFT-MD 7.13; Its parameters are shown in boldface in Table 1.
The $P(\rho)$ barotrope for Model DFT-MD 7.13 shown in
Figure~\ref{Pvsrho_gap}.  The corresponding $T(\rho)$ profile is shown
in Figure~\ref{TvsP_gap}.

\section{Compositional Perturbations to Equation of State} \label{comppert}

In order to derive general barotropes,
we must now evaluate the effects of (1) varying He concentration, and
(2) varying metallicity.  The barotrope shown in Figures
~\ref{Pvsrho_gap} and \ref{TvsP_gap} corresponds to an initial
He mass fraction $Y_0=0.245$ and metals mass fraction $Z_0=0$.  Since
this composition is a good initial approximation to the Jupiter
envelope, we use a perturbation approach to derive the effects of
compositional changes.  Let the reference barotrope for $Y_0=0.245$
and $Z_0=0$ be $\rho_0(P)$.  Although this barotrope is computed with
detailed DFT-MD simulations not assuming an ideal mixture of H and He,
to simplify this derivation we approximate it by an additive volume law,
$V_{\rm H-He}(P,T)=V_{\rm H}(P,T)+V_{\rm He}(P,T)$, valid for a noninteracting
mixture:
\begin{equation} \label{add_vol0}
{1 \over {\rho_0}} = {X_0 \over {\rho_H}} + {Y_0 \over {\rho_{He}}},
\end{equation}
where $X_0 = 1 - Y_0 =$ 0.755.  We now want to change the abundance of
helium to $Y$ and metals to $Z$.  We assume that the
temperature-pressure relation $T(P)$ is unchanged under perturbations
to the composition (i.e., the perturbing admixture is chemically and
thermodynamically inert).  With this assumption and the additive
volumes approximation, $V_{\rm H-He-Z}=V_{\rm H}+V_{\rm He}+V_Z$, the
perturbed density is given by,
\begin{equation} \label{add_vol1}
{1 \over {\rho}} = {{1 - Y - Z} \over {\rho_H}} + {Y \over {\rho_{He}}}
+ {Z \over {\rho_{Z}}},
\end{equation}
where $V_Z$ and $\rho_{Z}$ is the volume and density of the metals
component.  Rewriting Equations (\ref{add_vol0}) and (\ref{add_vol1}), we find
\begin{equation} \label{newrho}
{{\rho_0} \over {\rho}} = {{1-Y-Z} \over {1-Y_0}} + 
{{ZY_0+Y-Y_0} \over {1-Y_0}}{{\rho_0} \over {\rho_{He}}}
+ Z{{\rho_0} \over {\rho_{Z}}},
\end{equation}
with all densities evaluated for the reference $T(P)$. The same
equation is obtained if one starts from a fully interacting
hydrogen-helium equation of state and then perturbs the helium and
metals abundances.

For the composition in Jupiter's outer layers, at $P \phn \textless \phn 1$ Mbar, we
adopt abundances from Galileo Probe measurements \citep{Wong2004}.  In this region the
main contributors to $Z$ are the molecules ${\rm CH_4}$ and ${\rm NH_3}$,
and for the ${\rm H_2O}$ abundance we adopt the largest value measured
by the probe (rather than assuming a solar abundance for H$_2$O).
Neglecting other metals, we obtain $X =$ 0.7498, $Y =$ 0.2333, $Z \approx$ 0.0169,
for the presumed jovian composition at layers with $P \phn \textless \phn$ 1 Mbar.

\begin{figure}
\epsscale{1.0}

\plotone{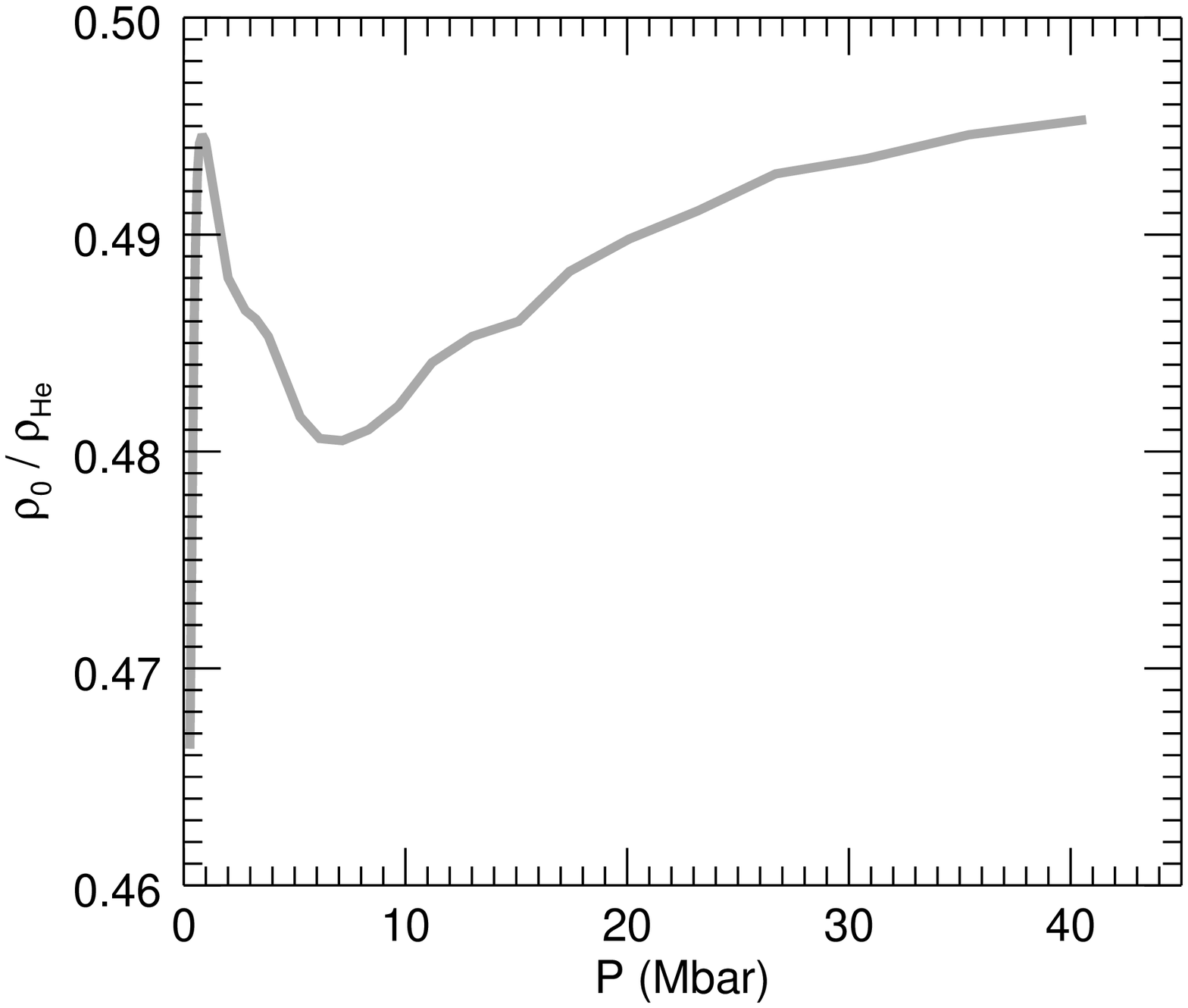}

\caption{
Results for ${{\rho_0} / {\rho_{He}}}$ , evaluated along a Jupiter barotrope,
to be inserted in 
Equation (\ref{newrho}).
}
\label{dratioHe}

\end{figure}

Using a DFT-MD equation of state for pure He (Militzer 2008) along the $T(P)$ shown in
Figure~\ref{TvsP_gap}, we obtain the density-pressure relation shown in Figure~\ref{dratioHe}.

\section{Equation of State of H$_2$O, CH$_4$, and NH$_3$} \label{hydrides}

Evaluation of the perturbation term $\rho_0 / \rho_Z$ in
Eq.(\ref{newrho}) is somewhat more complex because of the presence of
multiple molecular species, but need not be highly precise because the
contribution of this term is comparatively small.  We continue to
assume, for pressures above and below the He-immiscibility gap, that
the main contributors to the $Z$ mass fraction are the molecules ${\rm
  H_2O}$, ${\rm CH_4}$, and ${\rm NH_3}$, in solar proportions. Thus,
to evaluate $\rho_0 / \rho_Z$, it is necessary to evaluate the density
change of these molecular entities along the jovian $T(P)$.
We thus performed a number of DFT-MD simulations of
${\rm H_2O}$, ${\rm CH_4}$, and ${\rm NH_3}$ under such conditions.

All simulations were performed with the VASP code~\citep{vasp1} using
the PBE functional. Pseudopotentials of the projector-augmented wave
type~\citep{PAW} and a plane wave basis set cutoff of 1100~eV were
employed. The zone-average point,
$k=(\frac{1}{4},\frac{1}{4},\frac{1}{4})$, was used to sample the
Brillouin zone. A time step of 0.2 fs was used. Density, temperature,
and composition were prescribed in the simulations. After an initial
equilibration period, the pressure was derived by averaging over the MD
simulation.

We first benchmarked our simulations by comparing our results with the
shock wave measurements by~\citet{Nellis97} that compressed a mixture
of water, ammonia, and isopropanol (C$_3$H$_8$O) to 200 GPa. This
mixture, labeled ``synthetic Uranus'', was designed to resemble
the different planetary ices in the outer solar system. The
concentrations of the heavy nuclei (C:O=0.529, N:O=0.162) indeed
closely resemble solar proportions. However, the mixtures is somewhat
depleted in hydrogen (H:O=3.54) while one would expect a H:O ratio of
4.60 if one mixes H$_2$O, CH$_4$, and NH$_3$ in O:C:N proportions that
were used in the experiments. This difference prompted us to perform
two sets of simulations. First we studied a hydrogen-depleted
mixture, H:O:C:N=87:25:13:4, that closely resembles the ``synthetic
Uranus'' mixture within the size constraints of typical simulations
that accomodate between 100 and 200 atoms.

Our simulation results in Table~\ref{ice_table} show excellent
agreement with the experimental findings. It should be noted that if
we prescribe the central values for densities and temperature, that
were measured in the experiments, then our computed pressures were,
respectively, slightly higher and slightly lower than those reported
in the experiments.  However, if we adjusted the density and
temperatures in our simulations within the experimental 1 $\sigma$
uncertainties then our computed pressures fall within the experimental
error bars of the two available measurements. This provides another
example for DFT-MD simulations that closely reproduce experimental
findings~\citep{Knudson2012}.

\begin{figure}
\epsscale{1.0}
\plotone{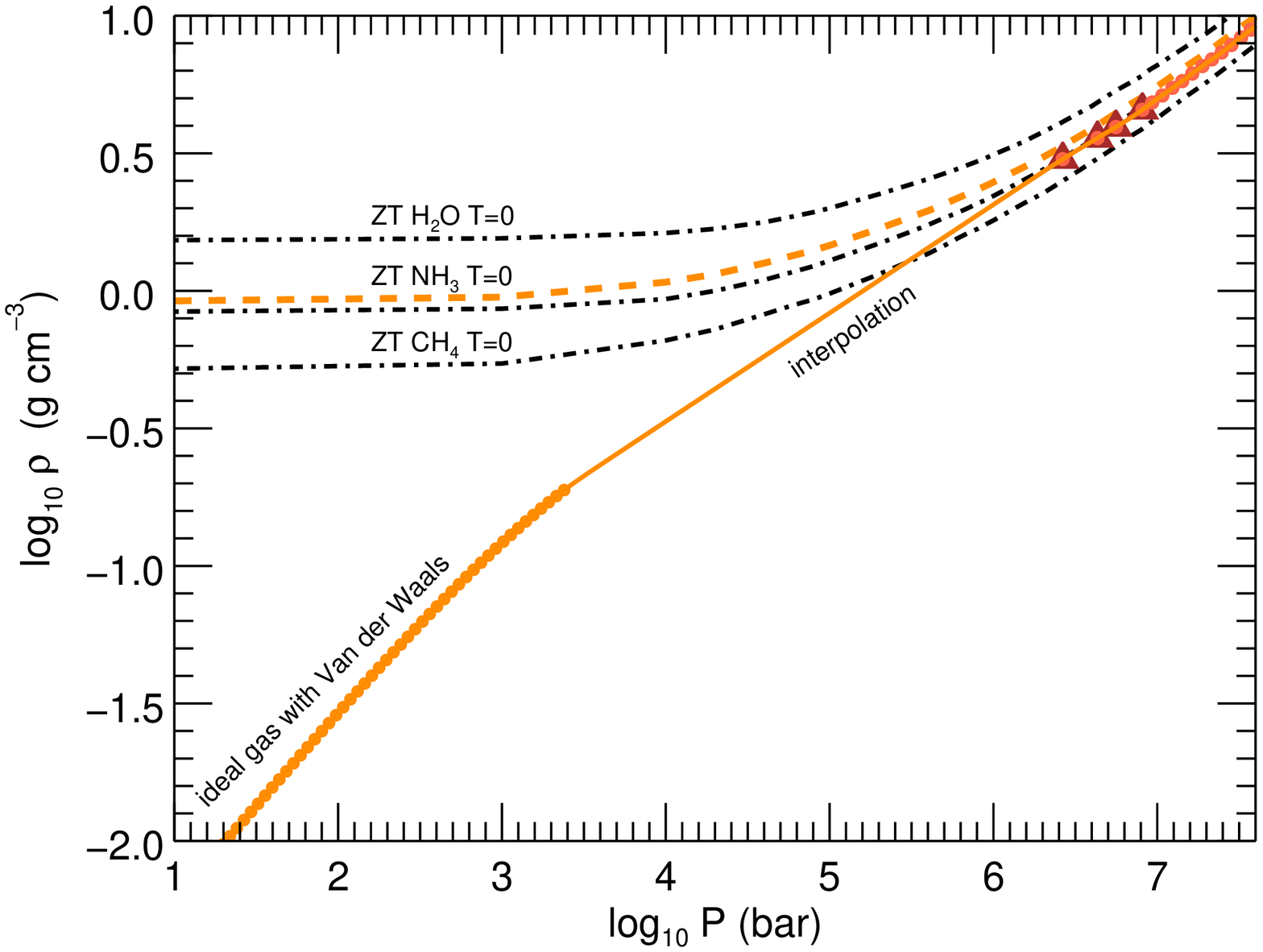}

\caption{
Procedure for determining the compression of a solar-proportions mixture of 
${\rm CH_4}$, ${\rm NH_3}$,and ${\rm H_2O}$ (the three
most important jovian hydrides) along a jovian $T(P)$ curve.
This relation is used to determine
$\rho_Z(P)$.  Van der Waals corrections for the three hydrides are computed using
data from \citet{Weast72}. }

\label{jupiter_hydrides}
\end{figure}

\begin{figure}
\epsscale{1.0}
\plotone{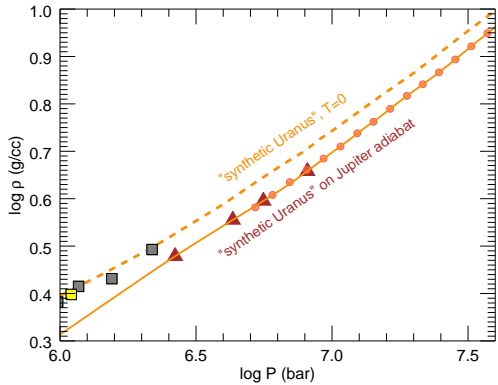}

\caption{
Expanded view of the high-pressure region of Figure~\ref{jupiter_hydrides}.
Brown triangles show results of our DFT-MD simulations for
a solar-proportions mixture of 
${\rm CH_4}$, ${\rm NH_3}$,and ${\rm H_2O}$ at four points on the jovian $T(P)$ curve.
These results overlap with results for a simple Mie-Gr{\"u}neisen
thermal perturbation (with a Gr{\"u}neisen $\gamma = 1$)
plus zero-temperature pressure, smaller red dots.
Squares show double-shock compression
points from Livermore gas gun experiments on ``synthetic Uranus'' carried
out by \citet{Nellis97}.
A temperature $T=4100 \pm 300$ K was measured for the data point at
1.1 Mbar, plotted as a yellow square.  A separate DFT-MD simulation agrees with this
data point to within the error bars, but is not used to calibrate our
$\rho_Z(P)$ curve.}
\label{jupiter_hydrides_hi}
\end{figure}

In Table~\ref{ice_table}, we also report results from simulations of a
H:O:C:N=99:21:12:3 mixtures that exactly represent the hydrogen
contents of a solar H$_2$O, CH$_4$, and NH$_3$ mixture. Because of the
higher hydrogen content, the density is lower than that of ``synthetic
Uranus'' when compared for the same $P$ and $T$. The simulation
results were incorporated into Figures~\ref{jupiter_hydrides} and
\ref{jupiter_hydrides_hi}.

Figure~\ref{jupiter_hydrides} shows calculations used to perform the
estimation of $\rho_0 / \rho_Z$.  In the low-pressure region of this figure, pressure-density
values for ${\rm CH_4}$, ${\rm NH_3}$, and ${\rm H_2O}$ are combined
assuming ideal mixing.  In the lower left-hand part of this figure,
orange dots show the ideal-gas partial pressure of an ideal mixture of
the three molecules along the jovian $T(P)$, with virial corrections
up to $P \sim 2$ Kbar.  Dash-dot curves at the top of the figure show
zero-temperature $\rho(P)$ relations calculated from
quantum-statistical models and tabulated in \citet{ZT1978}. The orange
dashed curve shows the resulting zero-temperature $\rho(P)$ relation
for a solar mixture of the three molecules.  Dots in the upper
right-hand corner of this figure show finite-temperature calculations
for pressures greater than a megabar; Figure~\ref{jupiter_hydrides_hi}
shows a zoom of this region, along with experimental data points for
``synthetic Uranus'' \citep{Nellis97}.

To construct $\rho_Z(P)$ in the gap between low pressure and high pressure,
we perform a linear interpolation in log-log space as indicated in
Figure~\ref{jupiter_hydrides}.

\begin{figure}
\epsscale{1.0}
\plotone{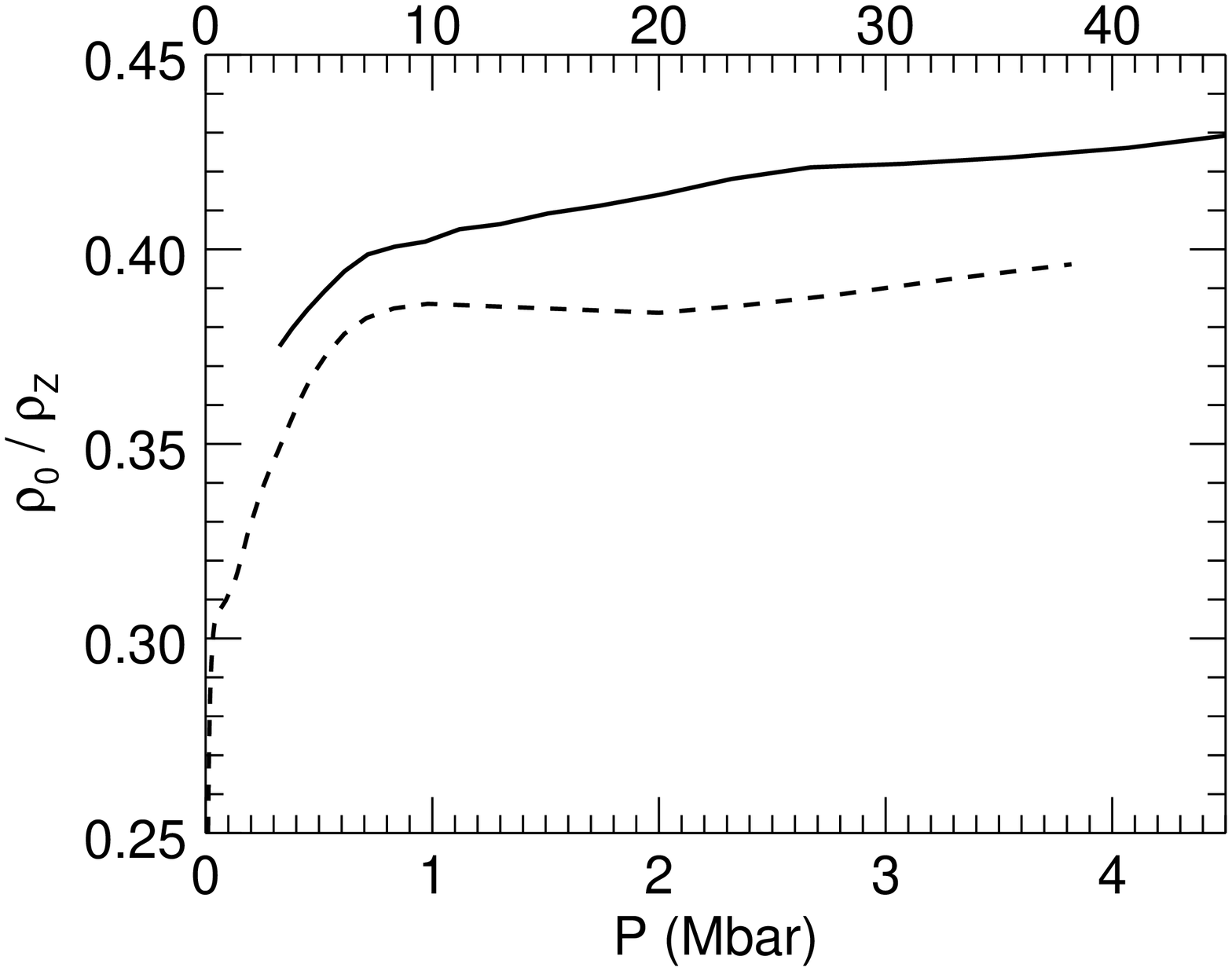}
\caption{
Results for ${{\rho_0} / {\rho_{Z}}}$ , evaluated along a Jupiter barotrope.
The dashed curve shows results for low pressures, spanning Jupiter's
``molecular'' layer (corresponding to the lower pressure axis).  The solid
curve (corresponding to the upper pressure axis)
shows results for pressures up to the core-mantle boundary
and slightly different composition, spanning Jupiter's
``metallic'' layer. These relations are inserted into Eq. \ref{newrho};
see Section \ref{hydrides} for details.}

\label{density_ratios_combined}

\end{figure}

Figure~\ref{density_ratios_combined} shows $\rho_0 / \rho_Z$ (dashed curve)
for assumed Galileo Probe composition (for pressures below 2 Mbar).
As a hypothesis to be tested by our preliminary Jupiter model,
we assume that all jovian layers at pressures less than $\sim 1$ Mbar have
the composition measured by the Galileo Probe, with a corresponding correction
to the density given by Equation (\ref{newrho}).  In this pressure range, we find from Figure~\ref{dratioHe} and Figure~\ref{density_ratios_combined} that $\rho_0 / \rho_{He} \approx$
0.48 and $\rho_0 / \rho_Z \approx$ 0.38, leading to $\rho_0 / \rho =$ 0.995.
The latter number is fortuitously close to unity because
the slightly lower Galileo Probe He abundance (relative to the DFT-MD simulations) is
almost compensated by the presence of metals.

Note that
${\rm H_2O}$ is depleted relative to ${\rm CH_4}$ and ${\rm NH_3}$
in the Galileo Probe data.  That is, Galileo Probe data show ${\rm H_2O}$
approaching a solar ratio to hydrogen-helium, while ${\rm CH_4}$ and ${\rm NH_3}$
are approximately three times their solar ratio to hydrogen-helium. 

In contrast, for solar proportions
of ${\rm CH_4}$:${\rm NH_3}$:${\rm H_2O}$ and for
$P \phn \textgreater \phn$ 1 Mbar, we have 
$\rho_0 / \rho_Z \phn \textgreater \phn$ 0.38, and
$\rho_0 / \rho_Z \approx$ 0.42 through the bulk of the jovian
envelope (Figure~\ref{density_ratios_combined}, solid curve).

For layers at pressures greater than 2.7 Mbar, we take the He and metals
abundances to be slightly higher than the protosolar values $Y =$ 0.2741 and
$Z =$  0.0149 \citep{Lodders03}.  Our DFT-MD equation of state, combined
with the constraints of Jupiter's total mass, volume, and $J_{2}$ and any reasonable
interior temperature distribution, does not imply a large increase of $Z$ above its
protosolar value, for otherwise the densities would be too large. 
Assuming $Z =$ 0.0246 in the deeper layers and
taking into account a slight He enrichment caused by
depletion in Jupiter's outer layers, we get $Y =$0.2788.  The
assumed value of $Z$ corresponds to abundances
of ${\rm CH_4}$ and ${\rm NH_3}$, relative to H, that are 4 times
protosolar.  Because of the much larger protosolar value of
${\rm H_2O}$ relative to H, a similar 4 times enhancement of this
molecule leads to larger $Z$ and hence interior densities,
and the resulting models would be outside the acceptable range.
We get $Z =$ 0.0246 if we take the enhancement of ${\rm H_2O}$ to be
2.4 times protosolar.

We insert this
value in Equation (\ref{newrho})
for the presumed jovian composition at layers with $P \textgreater$2 Mbar.
Then, over a pressure range corresponding to the bulk of the
jovian envelope, $2 \textless P  \textless 40$ Mbar,
we find from Figure~\ref{dratioHe} and Figure~\ref{density_ratios_combined}
that $\rho_0 / \rho_{He} \approx$
0.49 and $\rho_0 / \rho_Z \approx$ 0.42, leading to $\rho_0 / \rho =$ 0.959.

As is obvious from these results, and has long been known, the presence of metals
in Jupiter only affects the barotrope $\rho(P)$ at the level of a few percent.
Thus modeling the abundance and distribution of metals in
Jupiter by matching the planet's gravity data necessarily requires very accurate
(better than 1\%)
knowledge of $\rho_0(P)$.

\section{Jupiter Models}

\subsection{Spheroid Parameters and Code Function}

The version of the concentric maclaurin spheroid (CMS) code that we use
is designed to automatically calculate a mass distribution with a total
mass equal to Jupiter's mass, $M_J = 1.8986 \times 10^{30}$ g, and
an equatorial radius $a = 71492$ km.  The latter is the observed equatorial
radius of a layer at an average pressure of 1 bar, and the tabulated $J_{2n}$
are normalized to this radius.  We assume that Jupiter rotates as a solid
body with period \citep{Seid2007}
$P_{\rm rot} = 9^{\rm h} 55^{\rm m} 29.7^{\rm s} = 2 \pi / \omega$.
CMS theory is constructed to find a rotationally-distorted model for
the dimensionless small parameter
\begin{equation} \label{qdef}
q = {{\omega^2 a^3} \over {G M_J}},
\end{equation}
which to lowest order in $\omega^2$ is equivalent to $m$,
see Equation (\ref{mdef}), but is
more convenient as it can be directly computed from observed quantities.

Models are calculated with $N+1=511$ spheroids.  Using the notation of
\citet{Hu2013}, the dimensionless equatorial radii
of the spheroids $\lambda_i, i=0,...,N$ are specified as follows. By definition
$\lambda_0 \equiv$ 1 for the outermost spheroid (its equatorial radius $\equiv a$).  The
innermost spheroid surface is placed at $\lambda_{N} = 0.15$ (i.e., the core's
equatorial radius $= 0.15 \phn a$).  The choice of core radius is
somewhat arbitrary: the external zonal 
harmonic coefficients are sensitive to the total core mass but insensitive to its density.
Models have 170 spheroids equally spaced in $\lambda$ in the range
$0.15 \le \lambda \le 0.5$
and another 339 spheroids in the range
$0.5 \le \lambda \le 1 - \delta \lambda/2$. (where in this region
the spacing $\delta \lambda = 0.001477$,
or 105.6 km).
The outermost spheroid ($\lambda_0$) has zero density and is spaced
$\delta \lambda/2$ (or 53 km) above the next spheroid ($\lambda_1$).

We verified that zonal gravitational harmonic results were unaffected by the
details of the spheroid spacing, by carrying out
subsidiary calculations with spheroids
equally spaced from core to surface.  As shown in Figure~\ref{wt_fns}
for a typical Jupiter interior model, spheroids interior to
$\lambda \approx 0.5$ make no significant contribution to the
$J_n$.  Therefore we chose a closer spacing of spheroids exterior
to $\lambda = 0.5$, to improve accuracy.

As outlined in \citet{Hu2013}, two nested iterations are required to obtained
a converged rotationally-distorted model fitted to a given barotrope $P(\rho)$.
Before the iterations begin,
a provisional density distribution is specified, with each $i^{\rm th}$
spheroid having a constant density $\rho_i$.  For the specified $q$, the shape
and total potential of each $i^{\rm th}$ spheroid is then iteratively calculated
until relative changes between iterations fall below a specified tolerance, usually
$\sim 10^{-13}$.  Typically, this requires $\sim 30$ iterations.  
After satisfactory convergence, the total mass of the configuration $M_{\rm conf}$
is obtained by summing over all spheroids. 

An outer iteration loop (typically $\sim 50$ iterations)
 is performed to converge the model to the specified
barotrope $P(\rho)$.  As described by \citet{Hu2013}, using the $\rho_i$
and equipotential shapes from the converged inner loop, the
average pressure $P_i$ between the upper and
lower surface of each spheroid is calculated.
Then using the $P_i$, the barotrope relation is solved for each spheroid
to obtain new density values, $\rho_i = \rho(P_i)$.
The core spheroid is
not included in this procedure as it is assumed to be an incompressible
high-density region. See Section~\ref{improvement} for details on the convergence
of the iterations.

Define the renormalization
constant $\beta = M_J / M_{\rm conf}$.  After the latest outer
iteration, we renormalize all the $\rho_i$ by multiplying each
value (including the core) by the factor $\beta$.  These new
$\rho_i$ are then passed to the inner iteration loop, where the
spheroid shapes and corresponding external $J_{2n}$ are computed, and
then $M_{\rm conf}$ (which depends on the spheroid shapes) is
computed.  The resulting configuration is then passed back to the
outer loop.

The final result of the two iteration loops is a model with converged $J_{2n}$,
a mass and rescaled density of the incompressible core, and spheroids
$i=0,...,509$ fitted to the {\it scaled} prescribed barotrope $P = P(\beta \rho)$.
The model conforms precisely to the prescribed values of $q$, $a$, and
$M_J$.  The scaled barotrope $P = P(\beta \rho)$ corresponding to this model
is convenient for comparing with barotropes for various values of $Y$ and $Z$,
e.g. of the form of Equation (\ref{newrho}), in which the initial DFT-MD
simulations for $\rho_0(P)$ are rescaled by a (roughly constant) factor to
account for new values of $Y$ and $Z$.  Values of $\beta$ for each model are
used to obtain results for the model's metals content $Z$,
as entered in Table 1.  

Introduction of the renormalization constant $\beta$ provides a convenient method
for efficiently exploring the parameter space of jovian models, because, as
discussed in Section~\ref{hydrides}, to first approximation the density $\rho$ of a
perturbed mixture of H, He, and metals is related to the reference mixture
by the divisor  $\rho_0 / \rho$ which is nearly constant over a broad range of pressures.
Thus if $\beta \textless 1$, the overall metals content of the model is reduced with
respect to the assumed starting barotrope, and {\it vice versa}.

\subsection{Parameters of Barotrope and Core}

For the reader's assistance, Table 3 briefly defines a number of relevant parameters.

As discussed by \citet{MHVTB}, it is difficult to fit the
pre-{\it Juno} values of Jupiter's $J_{2n}$,
especially $J_4$,
with a constant-entropy, constant-composition barotrope and uniform rotation.  Although
the H-He DFT-MD equation of state has been updated since 2008, see
\citet{MH2013} and \citet{Militzer2013}, the difficulty remains.
For comparison purposes, we include at the end of Table 1
two interior models (denoted as SC) that we computed using
the same CMS procedure as the other models, but with
the older equation of state of \citet{SC95}.
These SC models are able to match the pre-{\it Juno} $J_{2}$ and
$J_4$ with vanishingly-small cores and tens of Earth masses
of metals in the envelope  (see Table 1).
Why are our DFT-MD models so different?  Although central temperatures for
DFT-MD and SC models are similar (see Table 1), it turns out that mid-envelope
temperatures for adiabatic DFT-MD models are considerably cooler.  This
behavior is a consequence of the depression of the adiabatic temperature gradient
associated with hydrogen metallization, as discussed by \citet{MH2013} and \citet{Militzer2013}.
Such behavior is not exhibited by the SC EOS and may not be
incorporated in the other recent Jupiter models.  Cooler
temperatures, as well as revisions to the pressure-density relation,
result in somewhat higher mass densities in the middle envelope, with respect to the other models.
It is this effect, in our models, that is primarily
responsible for considerably reduced envelope metallicity,
larger core mass, and increased  $|J_4|$. 

In 2008 we attempted to reduce the absolute value of $J_4$ by
hypothesizing a subrotating layer below Jupiter's observable
atmosphere, but this assumption is not supported by any realistic
circulation model.  It is possible to obtain a model which fits the
pre-{\it Juno} value of $J_4$ by instead introducing a chemical change
and corresponding extra density increase at layers around $P \phn \sim
\phn 1$ Mbar, but such models are not grounded in any fundamental
calculations of the thermodynamics of dense hydrogen plus impurities,
and are inconsistent with reasonable barotropes.  In this paper we
take a different approach.   We use the \citet{Mo2013}
  prediction for the pressure-temperature conditions of H/He
  immiscibility. Then we assume helium rain also introduces a
  composition change.  As discussed in Section 3, for $P \textless 1$
Mbar, we have $\rho_0 / \rho =$ 0.995, while for $P \textgreater 2.7$
Mbar, if one has four times solar (primordial) abundances of ${\rm
  CH_4}$, ${\rm NH_3}$, and $\sim 2.4$ times ${\rm H_2O}$, and no
other metals, as the composition at depth, one would have $\rho_0 /
\rho =$ 0.959.  These numbers suggest an expected extra $\sim 4\%$
density change resulting from the presence of a phase-separation
region and an increase of metallicity and helium to approximately
proto-solar values at deeper layers.  As we discuss in more detail
below, we need a much larger extra density change ($\sim 8\%$) to
obtain a DFT-MD model with $|J_4|$ reduced enough to agree with the
pre-{\it Juno} value.

To treat the expected extra density change, in the pressure
range between 1 and 2 Mbar we interpolate linearly in
$\log P$ and $\log \rho$ between the low-pressure barotrope with
$\rho = \rho_0(P) / 0.995$ and the high-pressure barotrope with
$\rho = \rho_0(P) / 0.959$, noting that $\rho_0(P)$ at 
$P \phn \textgreater \phn$ 2.7 Mbar lies on a higher-entropy adiabat
than the atmospheric adiabat.  Results for gravitational harmonic
coefficients of models are insensitive to the thickness of this narrow interpolation
region.  A CMS boundary could of course be placed at a discrete location
to exactly treat an actual density discontinuity, but the resulting change to the gravitational harmonic coefficients would be negligible.
 
The models presented in this paper are intended to correspond closely to
the theoretical behavior of hydrogen-helium mixtures and to properties of
the outer jovian layers as constrained by the Galileo Probe.  

The CMS method generates models that exactly fit the total jovian mass
and 1-bar equatorial radius.  We adjust the density (and thus the
mass) of the schematic central core of all models to obtain a match to
the pre-{\it Juno} observed value of $J_2$ given in Table 1, in the
expectation that a more precise post-{\it Juno} value will not differ
significantly from this number.  The other parameter beside the core
mass that is poorly constrained is the entropy of the deep adiabat,
which we vary from the Galileo Probe value $S =$ 7.08 through the
value that osculates the immiscibility boundary, $S =$ 7.20, on up to
(as an extreme case) $S =$ 7.24.  With increasing $S$, the thermal contribution to the deep
pressure increases, yielding lower density for a given pressure, thus
accommodating a slight increase in metallicity $Z$.  As we see from
Table 1, the predicted higher-order gravitational harmonic
coefficients vary from one model to the next at the level of $\sim
10^{-5}$ for $J_4$ (readily measurable by {\it Juno}), to $\sim
10^{-6}$ for $J_6$, to $\sim 10^{-8}$ for $J_8$.  The $J_{10}$ values
appear to have less value for discriminating interior structure, but
their near-constancy at a total level of $\sim 10^{-7}$ may be useful
as a reference for discerning the signature of nonhydrostatic effects
at a similar level, such as deep interior dynamics \citep{kas10}.

By increasing the density by an additional amount in the vicinity of
the He-immiscibility zone, it is possible to obtain a
match to Jupiter's pre-{\it Juno} $J_2$ and $J_4$ with a suitable model.
But, as noted by
\citet{MHVTB}, one does not have free rein in this process because Jupiter's
barotrope must correspond to a physically-plausible composition.  Because the
DFT-MD barotrope is generally denser than the corresponding barotrope that
one would compute using the theory of \citet{SC95}, in our DFT-MD models
very little enhancement of metals can be tolerated in Jupiter's envelope.

Most of our models are calculated
using (for $P \textless 1$ Mbar) the barotrope
$\rho = \rho_0(P,S = 7.08) / 0.995$ ,
corresponding to the Galileo Probe $T(P)$ and abundances, and the barotrope
$\rho = \rho_0(P,S) / 0.959$ for $P \textgreater 2.7$ Mbar,
corresponding to an adiabat with entropy $S \textgreater 7.08$,
(enhanced) protosolar helium abundance $Y=0.28$, and $Z=0.025$, corresponding
to Galileo-Probe enhancement of methane and ammonia and a lesser enhancement of
water,
but no presence of denser species such as magnesium-silicates.  During
the CMS calculations we linearly interpolate in $\log \rho$ vs. $\log P$ across the immiscibility region between 1 and 2.7 Mbar.
All models in Table 1 labeled DFT-MD $S$ (with no parenthesis) have the indicated compositions in
the molecular and metallic regions respectively.  As the deep $S$ increases, such models
show a modest increase in metallicity in the hydrogen-helium
envelope exterior to the dense core, as characterized by the parameter $M_Z$, the total mass
of metals in Earth masses.

Model DFT-MD 7.13 has $\beta=1.0000$, meaning that the input barotrope yields a
match to the total planetary mass without rescaling the densities.
A characteristic of DFT-MD 7.13 warrants discussion.  
This model has
Galileo Probe abundances of CH$_4$, NH$_3$, and H$_2$O {\it
  throughout} the molecular layer, and $4 \times$ solar abundances of
CH$_4$, and NH$_3$ in the metallic layer. The metallic layer
has $2.4 \times$ solar H$_2$O, more than in the molecular layer; a full
$4 \times$ solar H$_2$O enhancement would yield total densities which are too large
to fit the total mass of Jupiter.
As discussed in Section \ref{hydrides}, the assumed composition and
temperature profile results in a
reasonable $\rho(P)$ relation, which results in a reasonable planetary model.
However, acceptable $\rho(P)$ relations
only limit the possible range of temperature profiles
and metallicities but do not uniquely constrain them.

As alternatives, we investigated two variants of our preferred model, in which we imposed
equal metallicities in the molecular and metallic layers.  Model DFT-MD 7.13 (low-$Z$) has
artificially low $Z=0.004$ in both layers (although He abundance does increase from the Galileo
probe value to the protosolar value).  This unrealistic model has the largest $|J_4|$ and core mass
of the suite.  At the opposite extreme, Model 7.24 (equal-$Z$) has the same metallicity
$Z=0.027$ in both layers, and also has a relatively large $|J_4|$.  

All models shown in Table 1 have core mass adjusted
to give agreement to seven significant figures
with the observed value $J_2 = 14696.43 \times 10^{-6}$.  Two
of the models, DFT-MD 7.15(J4) and SC 7.15(J4), include an additional
density (and metallicity) increase across the immiscibility region between 1 and 2 Mbar,
adjusted to yield agreement with the pre-{\it Juno} observed values of $J_2$ and
$J_4 = -587.14 \times 10^{-6}$.  We note that uncertainties in
observed values in Table 1 are formal
error bars;  none of our models would be ruled out by these pre-{\it Juno}
measurements if the true error bars are $\sim 5$ times larger.
All of our models are close to the pre-{\it Juno}
observed value of $J_6$, but the agreement may be fortuitous.

\subsection{Comparison of Barotropes with Models}

\begin{figure}
\epsscale{1.0}
\plotone{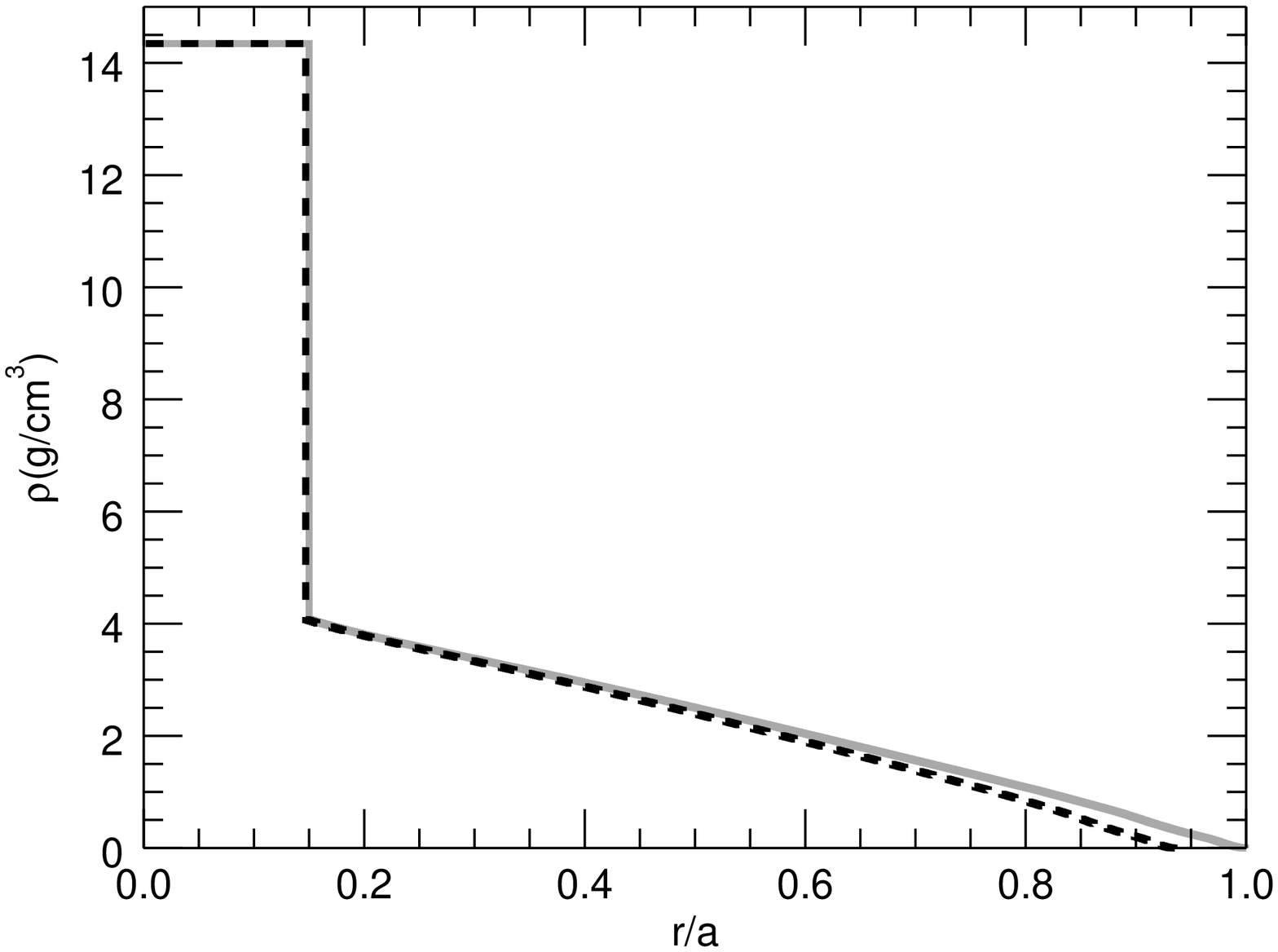}
\caption{
Equatorial (solid curve) and polar (dashed curve) density profiles.}
\label{rho_vs_r}
\end{figure}

Figure~\ref{rho_vs_r} shows a plot of polar and equatorial density
profiles for our preferred model DFT-MD 7.13.

\begin{figure}
\epsscale{1.0}
\plotone{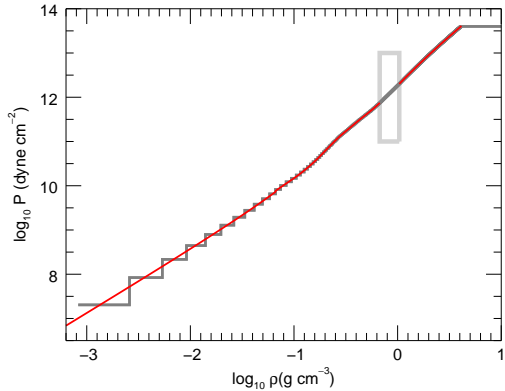}
\caption{
The grey stairstep shows converged CMS model DFT-MD 7.13. 
The light grey rectangle shows the region where He immiscibility
occurs and where the barotrope is interpolated to a higher-entropy
barotrope at higher pressure. The red curve is the
input barotrope for the assumed low-pressure and high-pressure
compositions.  
}
\label{P_vs_rho}
\end{figure}

\begin{figure}
\epsscale{1.0}
\plotone{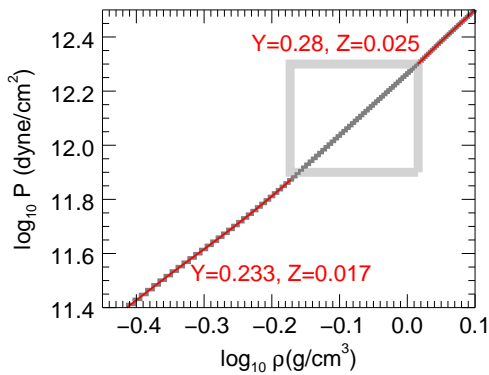}
\caption{
A close-up of the barotrope interpolation region for preferred CMS model DFT-MD 7.13. 
The red curve is the input
barotrope for the assumed compositions. }
\label{P_vs_rho_hi}
\end{figure}

\begin{figure}
\epsscale{1.0}
\plotone{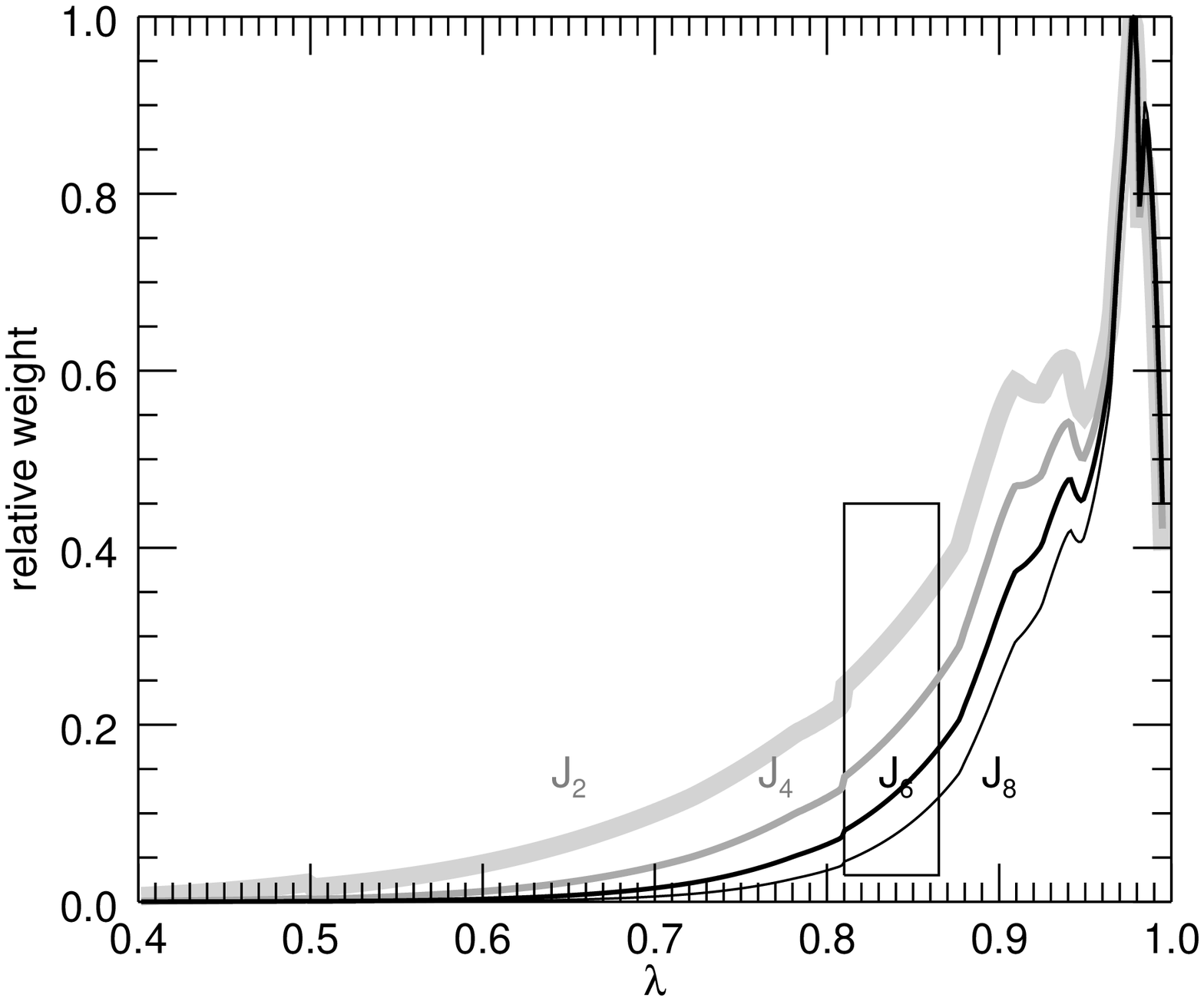}
\caption{
Relative contribution of spheroids to external gravitational zonal harmonic
coefficients,
for model DFT-MD 7.13.}
\label{wt_fns}
\end{figure}

Figure~\ref{P_vs_rho} plots the density vs. pressure profile for preferred
model DFT-MD 7.13 (grey stairstep), along with the input barotrope. 
Figure~\ref{P_vs_rho_hi} is a close-up of the high-pressure
region of Figure~\ref{P_vs_rho}.  The weighting functions
for contributions to the external zonal harmonic coefficients,
for the preferred model, are shown in
Figure~\ref{wt_fns}.

Model DFT-MD 7.15(J4) reduces $|J_4|$ to the observed value by
decreasing the barotrope's density at low pressures, and increasing the density at
high pressures.  However, densities in the outer region at pressures below 1 Mbar
then correspond to unphysical negative metallicity.  The entry for this model in Table 1
shows a total metals content $M_Z = 14.3 \phn M_E$ exterior to the dense core; this
value is the sum of $14.9 \phn M_E$ in the H-He envelope at pressures greater than
$\sim 1$ Mbar, and (unphysical) $-0.6 \phn M_E$ at lower pressures.  We are unable to
find a consistent DFT-MD Jupiter model that matches the observed
$J_2$ and $J_4$ values in Table 1.

\section{Moment of Inertia}

Jupiter's normalized moment of inertia NMoI $=C/Ma^2$ (where $C$ is the 
moment of inertia about the rotation axis) is in principle separately
measurable from the $J_{2n}$, and is a separate constraint on interior
structure. \citet{Helled2011} investigate models with fixed values of
$J_2$ and $J_4$ and conclude that a range of NMoI values between
0.2629 and 0.2645 can be found.
\citet{nettelmann2012} calculate a moment of inertia
but normalize it to the {\it mean} radius of the 1-bar
equipotential surface, a model-dependent quantity with a precision
limited to third order in their perturbative theory of figures.
However, their result is in reasonable agreement with values that we calculate below.
Since the nonperturbative approach
of our present investigation virtually eliminates any uncertainty in
the theoretical calculation of the $J_{2n}$, here we explore
the subject further as a guide to measurement requirements for the
{\it Juno} spacecraft.

Once a converged interior model is obtained, the NMoI is given exactly by
the expression
\begin{equation} \label{NMoI}
{C \over {Ma^2}} = {2 \over 5}{ {\Sigma_{j=0}^{N-1} \delta \rho_j
\int_0^1 d\mu \xi_j(\mu)^5} \over {\Sigma_{j=0}^{N-1} \delta \rho_j
\int_0^1 d\mu \xi_j(\mu)^3}} + {2 \over 3}J_2,
\end{equation}
in the notation of \citet{Hu2013}.

\begin{figure}
\epsscale{1.0}
\plotone{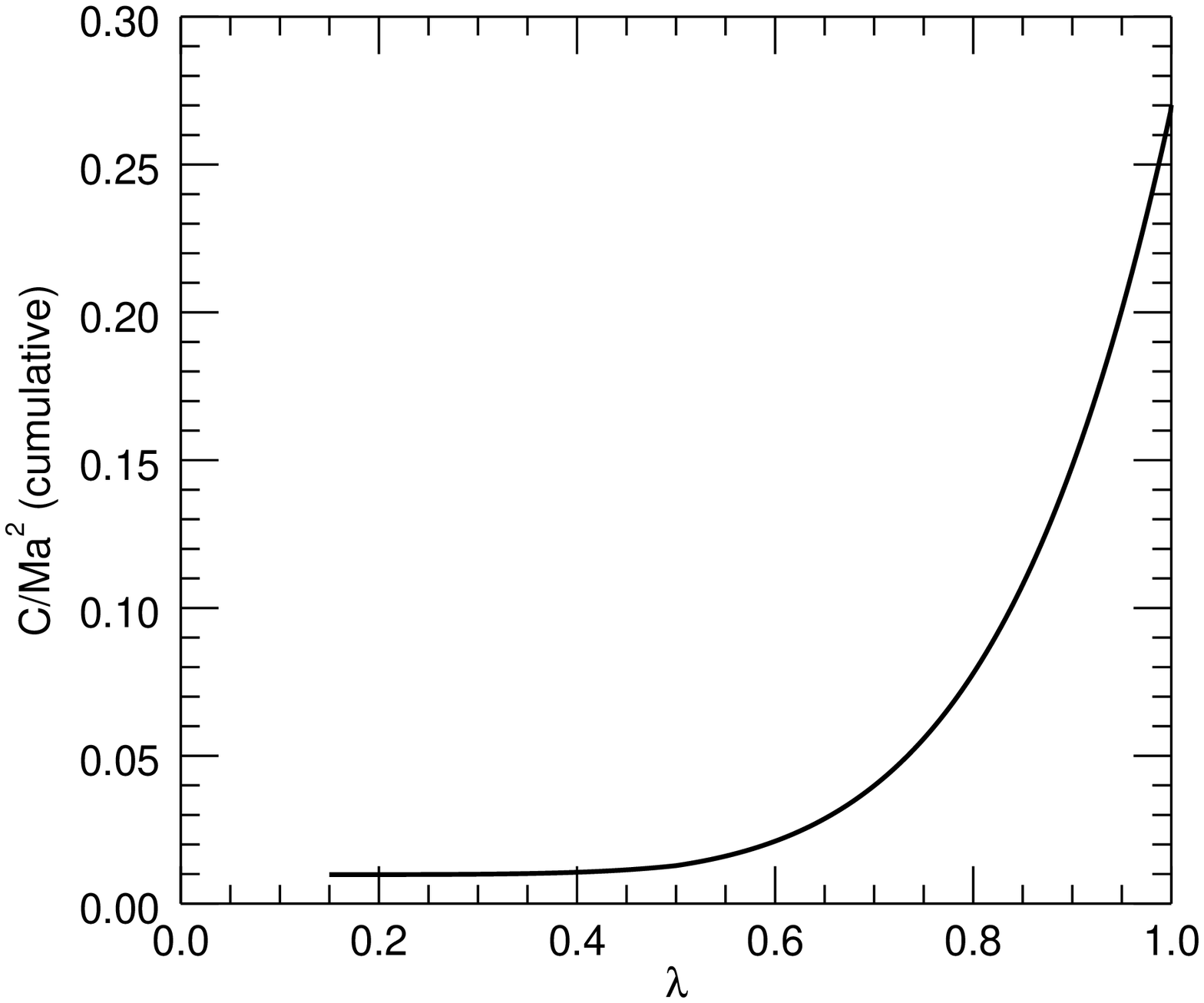}
\caption{
Cumulative value of $C/Ma^2$ for the preferred Jupiter model.  The final
point at $\lambda=1$ is the total value, $C/Ma^2 = 0.26389$.}
\label{coma2}
\end{figure}

Although Equation (\ref{NMoI}) resembles the Radau-Darwin relation in that it
seemingly relates the NMoI to $J_2$, it actually has no relationship
because Equation (\ref{NMoI})
shows that for a fixed $J_2$, an infinity of different CMS density
distributions could enter into the first term.  On the other hand, since each of
those CMS density distributions is required to yield the fixed $J_2$, the
range of variation of NMoI is in actuality quite restricted.  To illustrate
the point, in Figure~\ref{coma2} we show the cumulative value of the
NMoI as a function of the CMS radius $\lambda$, for preferred model
DFT-MD 7.13.  The cumulative value of ${C / {Ma^2}}$ is obtained by
partially summing the expression in Equation (\ref{NMoI}) from
the central CMS ($j=N-1$) out to a CMS with dimensionless equatorial radius
$\lambda$. 

\begin{figure}
\epsscale{1.0}
\plotone{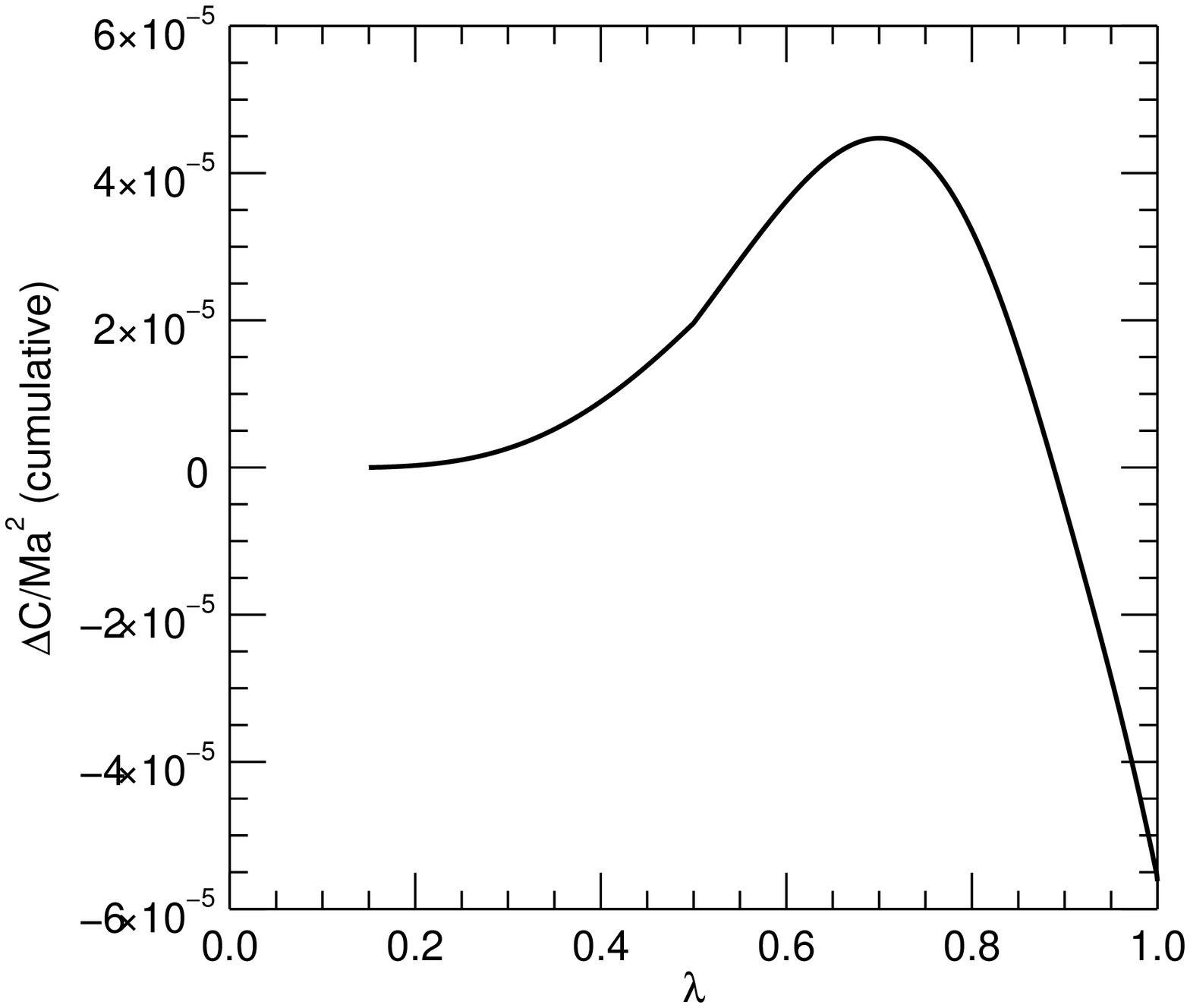}
\caption{
Difference in cumulative values of $C/Ma^2$ for the preferred Jupiter model
minus model with the SC equation of state. }
\label{Deltacoma2}
\end{figure}

To illustrate how details of interior structure affect the total NMoI,
Figure~\ref{Deltacoma2} shows the {\it difference} of the cumulative
values of $C/Ma^2$, for the preferred model minus model SC 7.15.   

\begin{figure}
\epsscale{1.0}
\plotone{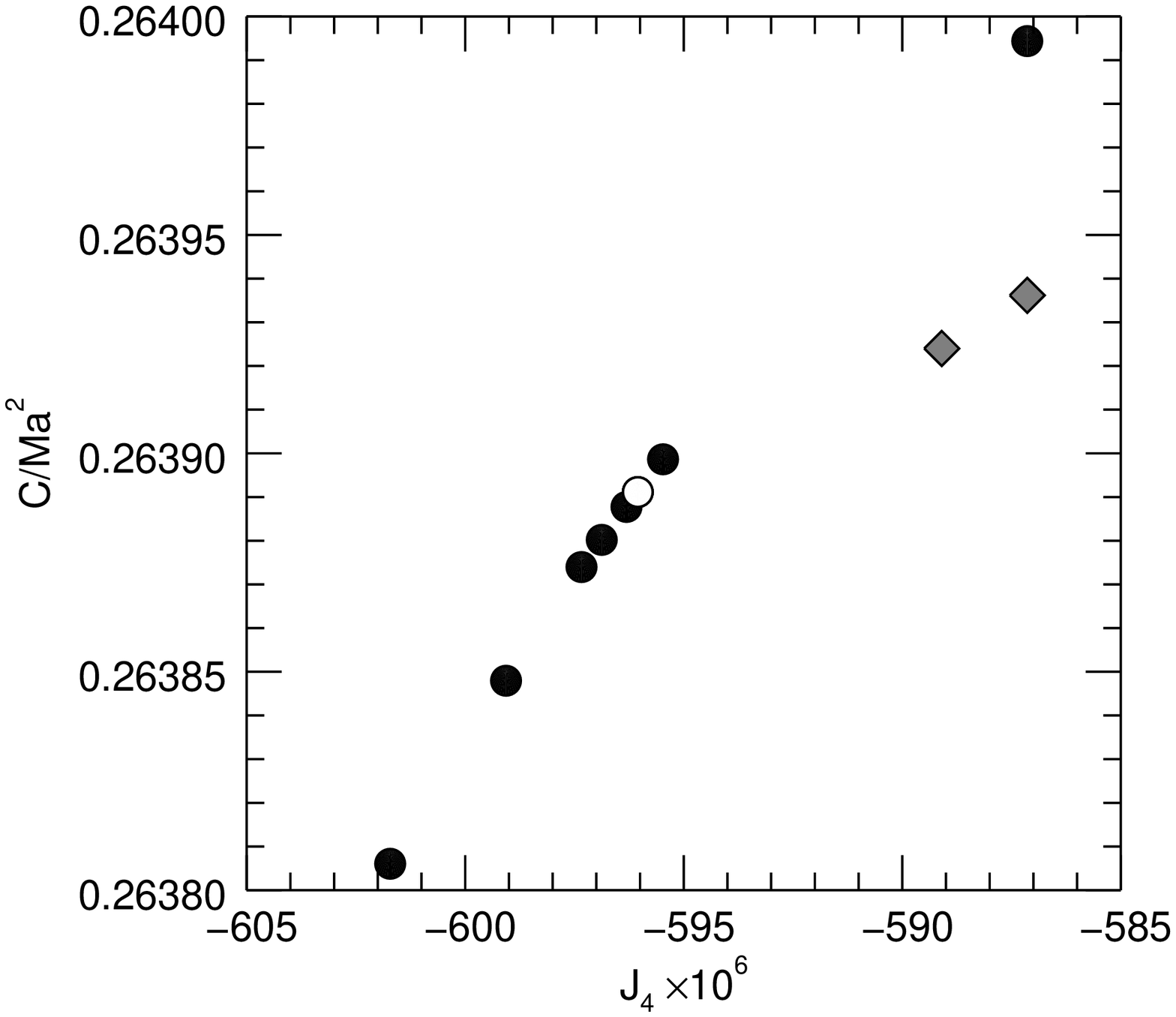}
\caption{
For the ten interior models of Table 1, all fixed to the observed
$J_2$, we plot the NMoI vs. $J_4$.  The open circle is the
preferred model.  The two diamonds to the right are the SC models.}
\label{coma2vsJ4}
\end{figure}

To truly discriminate between models with different barotropes,
it will be necessary to measure the NMoI to about five
significant figures, posing a difficult challenge to {\it Juno} or
other future investigations.
Figure~\ref{coma2vsJ4} illustrates the point.

We should point out that a measurement of Jupiter's NMoI would actually
be obtained from a measurement of the planet's spin angular momentum,
$J=C \omega$.  Thus if Jupiter were to rotate differentially on
cylinders with significant mass involved in the various rotation zones,
the tightly constrained values of NMoI that we find here might be broadened
to some extent.
It remains to be determined whether measurement of NMoI will prove to be
more of a constraint on the possibility of deep differential rotation, or
on the range of possible interior barotropes.

\section{Discussion and Conclusions}

The combination of the DFT-MD equation of state and observed
$J_{2n}$ already strongly limit the parameter space of acceptable pre-{\it Juno} models.

Our study has the following new features:
(a) We eliminate arbitrary density enhancements to fit the gravity field;
instead we utilize the H-He immiscibility phase boundary computed by
\citet{Mo2013} to bound the location and magnitude of a helium-related
compositional change;
(b) Our models incorporate the latest version of the DFT-MD
equation of state, replacing the widely-used SC EOS theory \citep{SC95};
(c) We utilize CMS theory for the first time to calculate high-order
zonal harmonic coefficients for realistic Jupiter models.

It is important to note that for fixed
$J_2$, the computed value of $|J_4|$ is sensitive to
the density in the region of Jupiter's metallic-hydrogen envelope where He
immiscibility is predicted.  One may force an agreement with the pre-{\it Juno}
value of $J_4$ given in Table 1 by imposing a density enhancement across the interpolation
region which is much larger than the $\sim 4\%$ implied by an increase in He
to the primordial value above $P \phn \textgreater \phn$ 2.7 Mbar.  However, when
this is done, conservation of mass leads to
 a model with (formally) negative metallicity in the low-pressure
outer envelope.

The new DFT-MD equation of state generally yields a very limited suite of
interior models of relatively low metallicity.
These models could be falsified by forthcoming {\it Juno} gravity
data.

In Jupiter model DFT-MD 7.13,
about 0.83 of the total mass is between the He-immiscibility region near
1 Mbar pressure and the core-mantle boundary.  So if $Z \sim 0.032$ in this region, the mass
of metals outside the core would comprise $\sim 10 M_E$, to be added to a core mass
$\sim 12 M_E$, for a total Jupiter metallicity $Z_{\rm global} \sim 0.07$. As
shown in Table 1, most of the other DFT-MD models have similar total
metallicities. In contrast, our models based on the SC EOS
(last two lines in Table 1) have
total metallicities that are about 60\% higher, in qualitative agreement with earlier results obtained by \citet{GGH97} and \citet{G99} that were also derived using
the SC equation of state.  The latter studies included the possibility
that Jupiter's core mass might be zero, and our independent SC models also
show very small core masses.  

The inferred large core masses of our DFT-MD models are
consistent with a core-nucleated scenario for the formation of Jupiter
\citep{Dangelo2014}.  The overall metallicity of Jupiter implied by
most of our models is roughly three times protosolar, implying that
about two-thirds of the volatile protosolar nebular complement to the
$\sim 12 M_E$ refractory core was not incorporated in primordial Jupiter.

In summary, we are able to derive Jupiter interior
models that match measured values of $J_2$, and sometimes $J_4$, and $J_6$, and are
consistent with predictions from published {\it ab initio} simulations
of hydrogen and helium, and additional results for different planetary
ices, H$_2$O, CH$_4$, and NH$_3$ that we report here. In our preferred
model, the heavy element abundance in the metallic layer is equivalent
to a three-fold solar concentration of all three ices. The preferred
value for the concentration in the molecular layer is slightly less
but consistent with the Galileo measurements.

Our preferred model has a massive core of 12 Earth masses which is
very similar to our earlier model~\citep{MHVTB}. When one uses the
semi-analytical equation of state (SC EOS) of \citet{SC95} instead of
our {\it ab initio} DFT-MD EOS, a much smaller core of 4 Earth masses
is predicted for the same model assumptions. This
illustrates how sensitively some model predictions depend on the
details of hydrogen-helium EOS.

Our Jupiter model is preliminary and intended for use as a reference for comparison
with experimental results from the {\it Juno} orbiter and other data sources.
New data will tell us how well the model works.

\acknowledgments

This work has been supported by NASA and NSF. 



\appendix

\section{Definitions for theory of figures}

The external potential
of a liquid planet in hydrostatic equilibrium rotating at a uniform rate $\omega$ is
usually expanded on Legendre polynomials $P_{2n}(\mu)$ as
\begin{equation} 
V(r,\mu)= {GM \over r}
\left[ 1 - \sum_{n=1}^{\infty} \left({a \over r} \right)^{2n} J_{2n} P_{2n}(\mu) \right] 
\end{equation}
where $G$ is the gravitational constant, $M$ the planet's mass, $a = 71492$ km
is the normalizing radius,
$\mu$ is the cosine of the angle from the rotation axis, and $r$ the radial distance
from the center of mass.  Pre-$\it {Juno}$ values of Jupiter's zonal
harmonic coefficients $J_{2n}$ are given in the first line of Table 1, and
are identical to values cited by \citet{MHVTB}.

It is expected that the $\it {Juno}$ gravity experiment will improve
the precision of the harmonic coefficients by at least two orders of magnitude and
measure the coefficients to degree 10 and possibly
beyond. Values of the $J_{2n}$ provide integral constraints on the
mass distribution within Jupiter, and can thus be used to constrain
interior models.  As discussed by \citet{HSKZ2014}, the basic
parameter that determines the magnitude of the $J_{2n}$ is the
dimensionless number $m$ (to lowest order, $m$ is the ratio of the
magnitude of the rotational acceleration to gravitational acceleration,
at the planet's equator),
\begin{equation} \label{mdef}
m = {{3 \omega^2} \over {4 \pi G \overline{\rho}}},
\end{equation}
where $\overline{\rho}$ is Jupiter's mean density.  \citet{ZT1978} show that one may
write
\begin{equation} \label{J2nsum}
J_{2n} = m^n \sum_{t=0}^{\infty} \Lambda_{2n}^{(t)} m^t,
\end{equation}
where the dimensionless response coefficients, $\Lambda_{2n}^{(t)}$, can be
obtained from the solution of a hierarchy of nonlinear perturbation
equations.  These response coefficients in turn depend on the equation of
state relating the pressure $P$ to the mass density $\rho$ at each point within the planet.
Provided that a barotropic relation $P(\rho)$ exists and that the planet is in hydrostatic
equilibrium, the perturbative potential-theory approach of \citet{ZT1978} can be used.  However,
$m \approx 0.08$ for Jupiter and $m \approx 0.14$ for Saturn, and the dimensionless coefficients
$\Lambda_{2n}^{(t)}$ do not decline rapidly with $n$ and $t$.  Replacing the infinite sum in
Equation (\ref{J2nsum}) with a finite sum up to, say $t \approx 9$ might suffice to determine the
measurable $J_{2n}$ to better than {\it Juno} precision, but would entail evaluation
of lengthy analytic expressions.  Instead, in this paper we use the more
straightforward non-perturbative concentric maclaurin
spheroid (CMS) theory of figures of \citet{Hu2013}.

\section{Numerical precision of CMS calculations}  \label{precision}

\begin{figure}
\epsscale{1.0}
\plotone{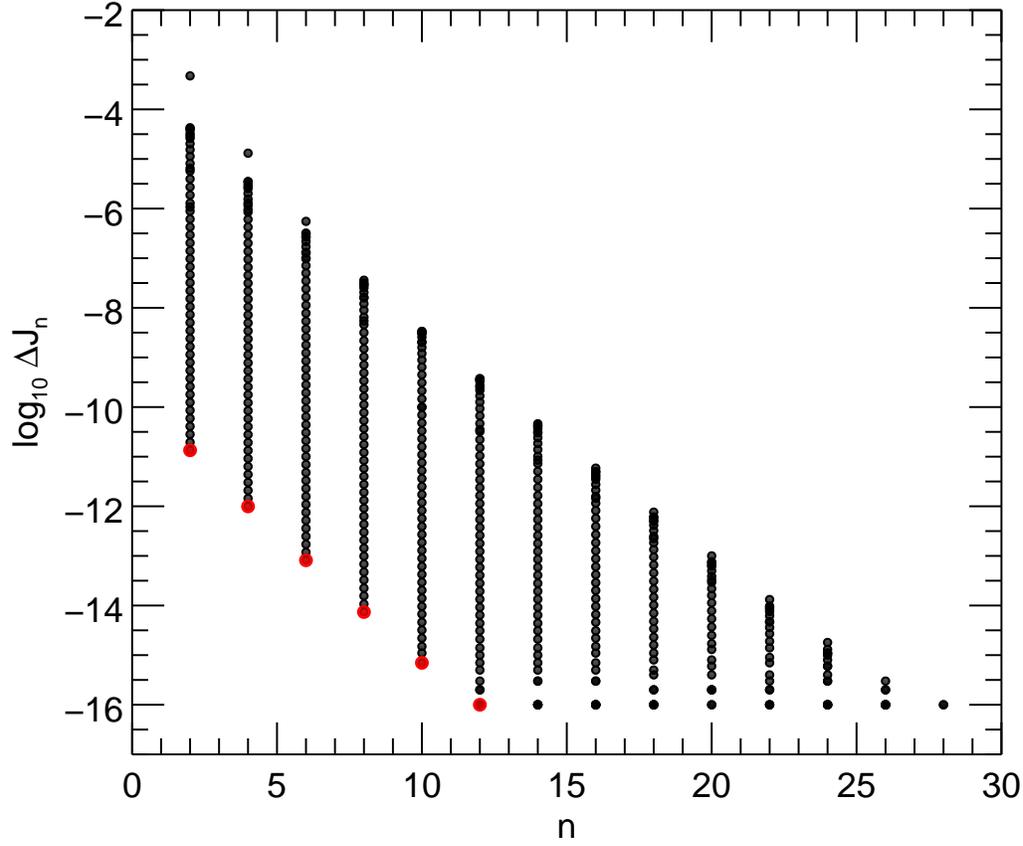}
\caption{
Improvement in the value of external zonal gravitational harmonic coefficients over 50 steps
in the outer iteration loop.  Here $\Delta J_n$ is the absolute value of
the change of $J_n$ from the previous iteration; the lowest points are
the values of the changes after the last iteration.
\label{improvement}}

\end{figure}

Figure~\ref{improvement} shows the improvement in the $J_{2n}$ values
for a typical model over 50 steps in the outer iteration loop.  After
50 iterations, the change in $J_{12}$ and higher degrees has fallen
below the computer's floating point precision.  The change in $J_2$
after 50 steps is at the level of $10^{-11}$, much smaller than the
precision with which it can be measured.

\begin{figure}
\epsscale{1.0}
\plotone{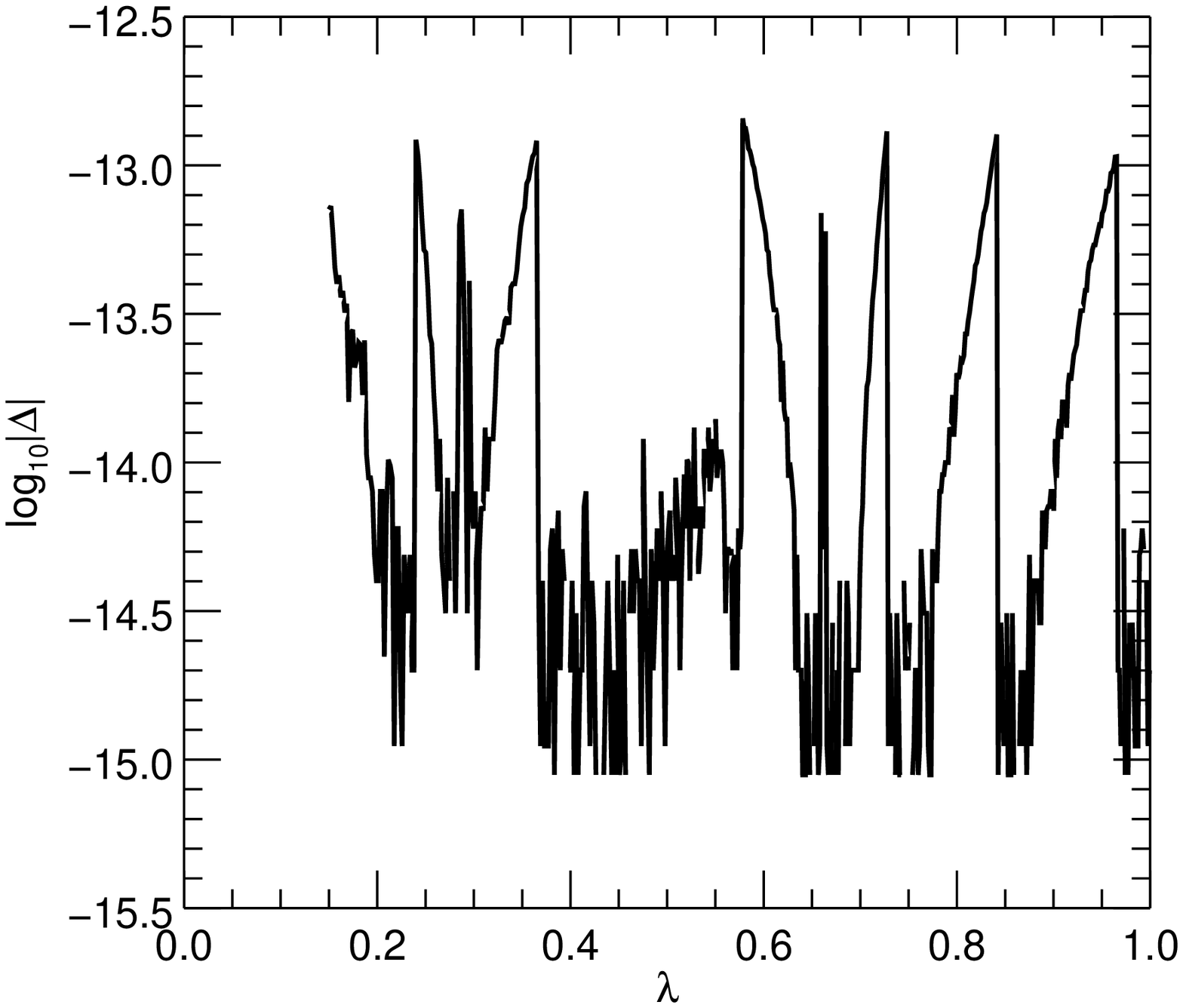}
\caption{
Here $\Delta$ is
the difference between the polar potential and the equatorial
potential (in units of $GM/a$), evaluated
after 30 inner-loop iterations within 50 outer-loop iterations.
\label{deltaplot}}
\end{figure}

Figure~\ref{deltaplot} shows the relative error in the CMS calculation
of the gravitational potential on the level surfaces of a converged
model using the audit-point method described in \citet{HSKZ2014}.

\eject





\clearpage

\begin{deluxetable}{lrrrrrrrrrr}
\rotate
\tablecolumns{11}
\tablewidth{0pc}
\tablecaption{Jupiter Zonal Harmonic Coefficients\tablenotemark{a} and Model Values
 (preferred model in boldface\tablenotemark{b})}
\tablehead{
\colhead{(all $J_n \times  10^6$)} & \colhead{$J_4$} & \colhead{$J_6$} & \colhead{$J_8$} & \colhead{$J_{10}$} 
& \colhead{$C / Ma^2$} &\colhead{$M_{\rm core}$} & \colhead{$M_{Z,{\rm molec.}}$} 
& \colhead{$M_{Z,{\rm metal.}}$} & \colhead{$Z_{\rm global}$} & \colhead{$T_{\rm CMB}$} \\ 
&&&&&&  \colhead{$(M_E)$}  &  \colhead{$(M_E)$} &  \colhead{$(M_E)$}  & & (K) }
\startdata
pre-Juno observed               & $-587.14         $  & $34.25         $  &  \nodata  &  \nodata  &  \nodata   &  \nodata &  \nodata &  \nodata      \\
\phn (JUP230)\tablenotemark{a}   & $        \pm 1.68$   & $      \pm 5.22$   &           &           &            &        &    &  & & \\
DFT-MD 7.24       & $-597.34$   & $35.30$   &  $-2.561$   &  0.212     & 0.26387   &  12.5     & 0.9 & 10.3 & 0.07 & 17600 \\
DFT-MD 7.24 {\footnotesize(equal-$Z$)} & $-599.07$   & $35.48$   &  $-2.579$   &  $0.214$   & 0.26385  &  13.1   & 1.1 & 7.5 & 0.07 & 17650 \\
DFT-MD 7.20       & $-596.88$   & $35.24$   &  $-2.556$   &  0.211     & 0.26388   &  12.3     & 0.8 & 9.9 & 0.07 & 17260 \\
DFT-MD 7.15       & $-596.31$   & $35.18$   &  $-2.549$   &  0.211     & 0.26389   &  12.2     & 0.7 & 9.2 & 0.07 & 16860 \\
DFT-MD 7.15 {\footnotesize($J_4$)} & $-587.14$   & $34.17$   &  $-2.450$   &  $0.201$   & 0.26399  &  $9.7$   & $-0.6$ & 14.9 & 0.08 & 16770 \\
\bf{DFT-MD 7.13}  & \bm{$-596.05$}& \bm{$35.15$}&  \bm{$-2.546$}&  \bm{$0.210$}& \bf{0.26389}&  \bm{$12.2$} & \bm{$0.7$} & \bm{$8.9$} & \bm{$0.07$} & \bm{$16670$} \\
DFT-MD 7.13 {\footnotesize(low-$Z$)} & $-601.72$   & $35.77$   &  $-2.608$   &  $0.217$   & 0.26381  &  14.0   & 0.2 & 1.1 & 0.05 & 16820 \\
DFT-MD 7.08       & $-595.47$   & $35.08$   &  $-2.539$   &  $0.210$    & 0.26390  &  $12.0$     & 0.6 & 8.3 & 0.07 & 16220 \\
SC 7.15           & $-589.10$   & $34.86$   &  $-2.556$   &  $0.214$   & 0.26392   &  $4.8$      & 3.5 & 28.2 & 0.11 & 18020 \\
SC 7.15 ($J_4$)   & $-587.14$   & $34.65$   &  $-2.534$   &  $0.212$   & 0.26394   &  $4.3$      & 3.2 & 29.3 & 0.12 & 17310 \\

\enddata
\tablenotetext{a}{Observed values are from R. A. Jacobson (2003), JUP230 orbit solution, with $J_2 = (14696.43 \pm 0.21)
\times  10^{-6}$.  All theoretical models match $J_2 = 14696.43 \times  10^{-6}$ to seven significant figures.}
\tablenotetext{b}{$\sim 0.83$ of the total mass is in the metallic layer, i.e. between $\sim 2$ Mbar and the CMB at $\sim 40$ Mbar.}

\end{deluxetable}

\begin{deluxetable}{lllllll}
\tablecolumns{7}
\tablewidth{0pc}
\tablecaption{Comparison of shock wave measurements by \citet{Nellis97} and our 
{\it ab initio} simulations that used two compositions (a) H:O:C:N=87:25:13:4 and (b) H:O:C:N=99:21:12:3.}
\tablehead{
\colhead{Method}  & \colhead{H:O} & \colhead{C:O} & \colhead{N:O} & \colhead{$\rho$ (g$\,$cm$^{-3}$)}& \colhead{$T$ (K)}   & \colhead{$P$ (GPa)} }
\startdata                                                                                                                 
Experiment        &  3.54         & 0.529         & 0.162         & 2.044 $\pm$ 0.005                & ~3220 $\pm$ 200      & 49.9 $\pm$ 0.5      \\
Simulation$^{(a)}$&  3.48         & 0.520         & 0.160         & 2.044                            & ~3220                & 52.17$\pm$ 0.17     \\
Simulation$^{(a)}$&  3.48         & 0.520         & 0.160         & 2.039                            & ~3020                & 50.17$\pm$ 0.30     \\
\hline
Experiment        &  3.54         & 0.529         & 0.162         & 2.45~ $\pm$ 0.13                 & ~4100 $\pm$ 300      & 110~~~ $\pm$ 4        \\
Simulation$^{(a)}$&  3.48         & 0.520         & 0.160         & 2.450                            & ~4100                & ~96.34 $\pm$ 0.42\\
Simulation$^{(a)}$&  3.48         & 0.520         & 0.160         & 2.580                            & ~4400                & 114.47 $\pm$ 0.35\\
\hline
Simulation$^{(b)}$&  4.71         & 0.571         & 0.143         &  2.353  & ~4100 & 117.90  $\pm$ 0.32 \\
Simulation$^{(b)}$&  4.71         & 0.571         & 0.143         &  2.262  & ~4100 & 105.37  $\pm$ 0.25 \\
Simulation$^{(b)}$&  4.71         & 0.571         & 0.143         &  3.011  & ~7000 & 264.92  $\pm$ 0.48 \\
Simulation$^{(b)}$&  4.71         & 0.571         & 0.143         &  3.592  & ~8000 & 431.78  $\pm$ 0.34 \\
Simulation$^{(b)}$&  4.71         & 0.571         & 0.143         &  3.940  & ~9000 & 559.32  $\pm$ 0.42 \\
Simulation$^{(b)}$&  4.71         & 0.571         & 0.143         &  4.550  & 10000 & 811.62  $\pm$ 0.56 \\
\enddata
\label{ice_table}
\end{deluxetable}

\begin{deluxetable}{ll}
\tablecolumns{2}
\tablewidth{0pc}
\tablecaption{Definitions of some parameters used in this paper}
\tablehead{
\colhead{Parameter(s)}  & \colhead{Definition}  }
\startdata                                                                                                                 
$X_0$, $Y_0$    &  mass fractions of H and He in DFT-MD simulations; see Equation (\ref{add_vol0})     \\
$\rho_0$    &  mass density of H-He mixture in DFT-MD simulations, for given $P$ and $T$     \\
$X$, $Y$, $Z$    &  perturbed mass fractions of H, He, and metals; see Equation (\ref{add_vol1})     \\
$Z_{\rm global}$ & total mass fraction of ``metals'' in Jupiter (including dense core)     \\
$\rho_0/\rho$    &  ratio of mass density for reference barotrope, Equation (\ref{add_vol0}), to  \\
                         & \phn mass density with perturbed $X$, $Y$, $Z$    \\
$\rho_0/\rho_{He}$    &  ratio of mass density for reference barotrope, Equation (\ref{add_vol0}), to  \\
                                  & \phn mass density of pure He at same $P$ and $T$       \\
$\rho_0/\rho_{Z}$    &  ratio of mass density for reference barotrope, Equation (\ref{add_vol0}), to \\
                                & \phn mass density of a pure ``metals'' mixture at same $P$ and $T$       \\
$M_E$ &  mass of the Earth    \\
$M_{Z,{\rm molec.}}$ &  total mass of ${\rm CH}_4+{\rm NH}_3+{\rm H}_2{\rm O}$ in Jupiter's molecular layer; see Table 1    \\
$M_{Z,{\rm metal.}}$ &  total mass of ${\rm CH}_4+{\rm NH}_3+{\rm H}_2{\rm O}$ in Jupiter's metallic layer; see Table 1    \\
$T_{\rm CMB}$    &  temperature at the core-mantle boundary, generally at $P \approx$ 40 Mbar;  \\
                & \phn  see Table 1    \\
$\beta$    &  dimensionless factor applied to prescribed barotrope $P=P(\beta \rho)$ to yield  \\
                & \phn  exact Jupiter mass; equivalent to compositional perturbation    \\

\enddata

\label{definitions_table}
\end{deluxetable}

\end{document}